\documentclass{article}

\PassOptionsToPackage{numbers, compress}{natbib}
\usepackage[preprint]{neurips_2025}
\usepackage{amsmath}

\usepackage[utf8]{inputenc} 
\usepackage[T1]{fontenc}    
\usepackage{hyperref}       
\usepackage{url}            
\usepackage{booktabs}       
\usepackage{amsfonts}       
\usepackage{nicefrac}       
\usepackage{microtype}      
\usepackage{xcolor}         
\usepackage{graphicx}
\usepackage{subcaption} 

\newcommand{\floor}[1]{\left\lfloor #1 \right\rfloor}

\title{Active Speech Enhancement: Active Speech Denoising Decliping and Deveraberation}

\author{
  Ofir Yaish \\
  School of Electrical and Computer Engineering \\
  Ben-Gurion University of the Negev \\
  \texttt{yaishof@post.bgu.ac.il} \\
  \And
  Yehuda Mishaly \\
  Blavatnik School of Computer Science \\
  Tel Aviv University \\
  \texttt{mishaly1@mail.tau.ac.il} \\
  \And
  Eliya Nachmani\thanks{Corresponding author.} \\
  School of Electrical and Computer Engineering \\
  Ben-Gurion University of the Negev \\
  \texttt{eliyanac@bgu.ac.il} \\
}

\begin{document}
\maketitle

\begin{abstract}
We introduce a new paradigm for active sound modification: \textit{Active Speech Enhancement} (ASE). While Active Noise Cancellation (ANC) algorithms focus on suppressing external interference, ASE goes further by actively shaping the speech signal—both attenuating unwanted noise components and amplifying speech-relevant frequencies—to improve intelligibility and perceptual quality. To enable this, we propose a novel Transformer-Mamba-based architecture, along with a task-specific loss function designed to jointly optimize interference suppression and signal enrichment. Our method outperforms existing baselines across multiple speech processing tasks—including denoising, dereverberation, and declipping—demonstrating the effectiveness of active, targeted modulation in challenging acoustic environments.
\end{abstract}

\section{Introduction}
Speech enhancement and noise control are fundamental audio processing tasks. Speech enhancement aims to improve the perceptual quality and intelligibility of speech signals by mitigating degradations such as background noise, distortion, clipping, and reverberation. Traditional approaches—including spectral subtraction, Wiener filtering, and statistical model–based methods—have achieved varying degrees of success but often falter in highly non-stationary noise environments~\citep{boll2003suppression, lim1978all, paliwal2012speech}. Recent advances in deep learning have, however, yielded state-of-the-art performance: convolutional neural networks (CNNs)~\citep{pascual2017segan, rethage2018wavenet, pandey2018new}, recurrent neural networks (RNNs)~\citep{hu2020dccrn}, generative adversarial networks (GANs)~\citep{fu2019metricgan, Fu_2021, kim2021multi, shin2023metricgan, shetu2025gan}, Transformers~\citep{Wang_2021, de_Oliveira_2022, zhang2022cross, cao2022cmgan, ye2023improved, Zhang_2024}, and diffusion models~\citep{guimares2025ditse, lu2022conditional, welker2022speech, Richter_2023, Lemercier_2023, tai2023revisiting, Ayilo_2024} demonstrate exceptional results on benchmarks for denoising, dereverberation, and declipping.

Active noise cancellation takes a complementary approach by generating an anti‐noise signal to destructively interfere with unwanted background noise. Pioneering work dating back to Lueg’s first patent in 1936 introduced the concept of adaptive feedforward ANC~\citep{lueg1936process}, which was later refined through advances in adaptive filtering (e.g., LMS, FxLMS), multi‐channel algorithms, and applications in headphones and enclosure systems~\citep{nelson1991active, fuller1996active, hansen1997active, kuo1999active, zhang2021deep, park2023had, mostafavi2023deep, cha2023dnoisenet, singh2024enhancing, pike2023generalized, mishaly2025deepasc}. While ANC excels at suppressing predictable or narrowband noise, it does not actively modify the speech content itself.

In this paper, we propose a new paradigm—\textit{Active Speech Enhancement}—that unifies the goals of speech enhancement and active noise control. Unlike conventional ANC, which solely targets noise suppression, \textit{Active Speech Enhancement} (ASE) actively shapes the speech signal by simultaneously attenuating interfering components and amplifying speech‐related frequencies. This dual‐action approach not only reduces the noise but also emphasizes the speech and improves perceptual quality under challenging acoustic conditions. We make the following key contributions:
\begin{itemize}
    \item We formalize the ASE task and define appropriate evaluation metrics that capture noise suppression and speech enhancement objectives.
    \item We introduce a novel Transformer-Mamba-based model that learns to generate an active modification signal, leveraging self‐attention to capture long‐range dependencies in time–frequency representations.
    \item Joint Suppression–Enrichment Loss. We design a task‐specific loss function that balances interference removal and signal enrichment, combining spectral, perceptual, and adversarial objectives to drive the model toward optimal ASE performance.
    \item Comprehensive Evaluation. Our method outperforms baselines across multiple tasks, including speech denoising, dereverberation, and declipping, achieving significant gains in standard metrics such as PESQ~\citep{rix2001perceptual}.
\end{itemize}

A demo page is available at \href{https://ofiryaish.github.io/ASE-TM-Demo/}{our project demo page}.

\section{Related Work}

\subsection{Speech Enhancement}

Recent advances in deep learning have yielded substantial improvements in speech enhancement. Pandey and Wang~\citep{pandey2018new} proposed a CNN–based autoencoder that applies convolutions directly to the raw waveform while computing the loss in the frequency domain. SEGAN~\citep{pascual2017segan}, introduced by Pascual \emph{et al.}, employs strided convolutional layers in a generative adversarial framework. Rethage \emph{et al.}~\citep{rethage2018wavenet} developed a WaveNet‐inspired model that predicts multiple waveform samples per step to reduce computational cost. Recurrent architectures have also been explored. Hu \emph{et al.}~\citep{hu2020dccrn} presented DCCRN, which integrates complex‐valued convolutional and recurrent layers to process spectrogram inputs. A real-time causal model based on an encoder–decoder with skip connections was proposed by Défossez \emph{et al.}~\citep{defossez2020real}, operating directly on the raw waveform and optimized in both time and frequency domains.

MetricGAN~\citep{fu2019metricgan} and its successor MetricGAN+~\citep{Fu_2021} incorporate evaluation metrics such as PESQ (Perceptual Evaluation of Speech Quality)~\citep{rix2001perceptual} and STOI (Short‐Time Objective Intelligibility)~\citep{kim2021multi} into the adversarial loss. Kim \emph{et al.}~\citep{kim2021multi} further enhance this approach by introducing a multiscale discriminator operating at different sampling rates alongside a generator that processes speech at multiple resolutions.

Transformer‐based models have recently gained prominence. Wang \emph{et al.}~\citep{Wang_2021} proposed a two‐stage transformer network (TSTNN) that outperforms earlier time‐ and frequency‐domain methods. CMGAN~\citep{cao2022cmgan} adapts the Conformer backbone~\citep{gulati2020conformerconvolutionaugmentedtransformerspeech} for enhancement, and de Oliveira \emph{et al.}~\citep{de_Oliveira_2022} replace the learned encoder of SepFormer~\citep{subakan2021attentionneedspeechseparation} with long‐frame STFT (Short-Time Fourier Transform) inputs, reducing sequence length and lowering computational cost without compromising perceptual quality.

More recently, diffusion‐based approaches have emerged as a powerful generative paradigm. Lu \emph{et al.}~\citep{lu2022conditional} introduced a conditional diffusion probabilistic model that learns a parameterized reverse diffusion process conditioned on the noisy input. Welker \emph{et al.}~\citep{welker2022speech} extended score‐based models to the complex STFT domain, learning the gradient of the log‐density of clean speech coefficients. Richter \emph{et al.}~\citep{Richter_2023} formulate enhancement as a stochastic differential equation, initializing reverse diffusion from a mixture of noisy speech and Gaussian noise and achieving high‐quality reconstructions in only 30 steps. Lemercier \emph{et al.}~\citep{Lemercier_2023} propose a \emph{stochastic regeneration} method that leverages an initial predictive‐model estimate to guide a reduced‐step diffusion process, mitigating artifacts and reducing computational cost by an order of magnitude while maintaining enhancement quality.

\subsection{Active Audio Cancellation}
Recently, deep learning approaches have demonstrated remarkable results in ANC algorithms. Zhang et al.~\citep{zhang2021deep} introduced DeepANC, which employs a convolutional long short-term memory (Conv-LSTM) network to jointly estimate amplitude and phase responses from microphone signals. Subsequently, attention-driven ANC frameworks integrating attentive recurrent networks were proposed to enable real-time adaptation and low-latency operation~\citep{zhang2022attentive}.

A selective fixed-filter ANC (SFANC) framework was developed to leverage a two-dimensional CNN for optimal control-filter selection on a mobile co-processor and a lightweight one-dimensional CNN for time-domain noise classification, yielding superior attenuation of real-world non-stationary headphone noise~\citep{shi2022selective}. Luo et al.~\citep{luo2022hybrid} proposed a hybrid SFANC–FxNLMS that first applies a similar approach as SFANC for each noise frame and then applies the FxNLMS algorithm for real-time coefficient adaptation, thereby combining the rapid response of SFANC with the low steady-state error and robustness of adaptive optimization. Heuristic algorithms—such as bee colony optimization~\citep{ren2022improved} and genetic algorithms~\citep{zhou2023genetic}—have been explored to avoid gradient-based learning.

Other studies have applied recurrent convolutional networks~\citep{park2023had, mostafavi2023deep, cha2023dnoisenet} and fully connected neural networks~\citep{pike2023generalized} to ANC. Autoencoder-based encodings have been used to extract latent features for improved robustness~\citep{singh2024enhancing}. Efforts in SFANC have extended to synthesizing optimized filter banks via unsupervised methods~\citep{luo2024unsupervised}, while advancements in multichannel setups continue to leverage spatial diversity through deep controllers~\citep{shi2024behind}. Multichannel configurations have been further enhanced by refined deep controllers that learn inter-channel relationships for improved noise attenuation~\citep{zhang2023deep, antonanzas2023remote, xiao2023spatially, zhang2023time, shi2023multichannel}, and attention-driven frameworks have been investigated for low-latency operation~\citep{zhang2023low}.

\section{Background}
\begin{figure}[t]
  \centering
  \begin{subfigure}[t]{0.49\textwidth}
    \centering
    \includegraphics[width=\textwidth]{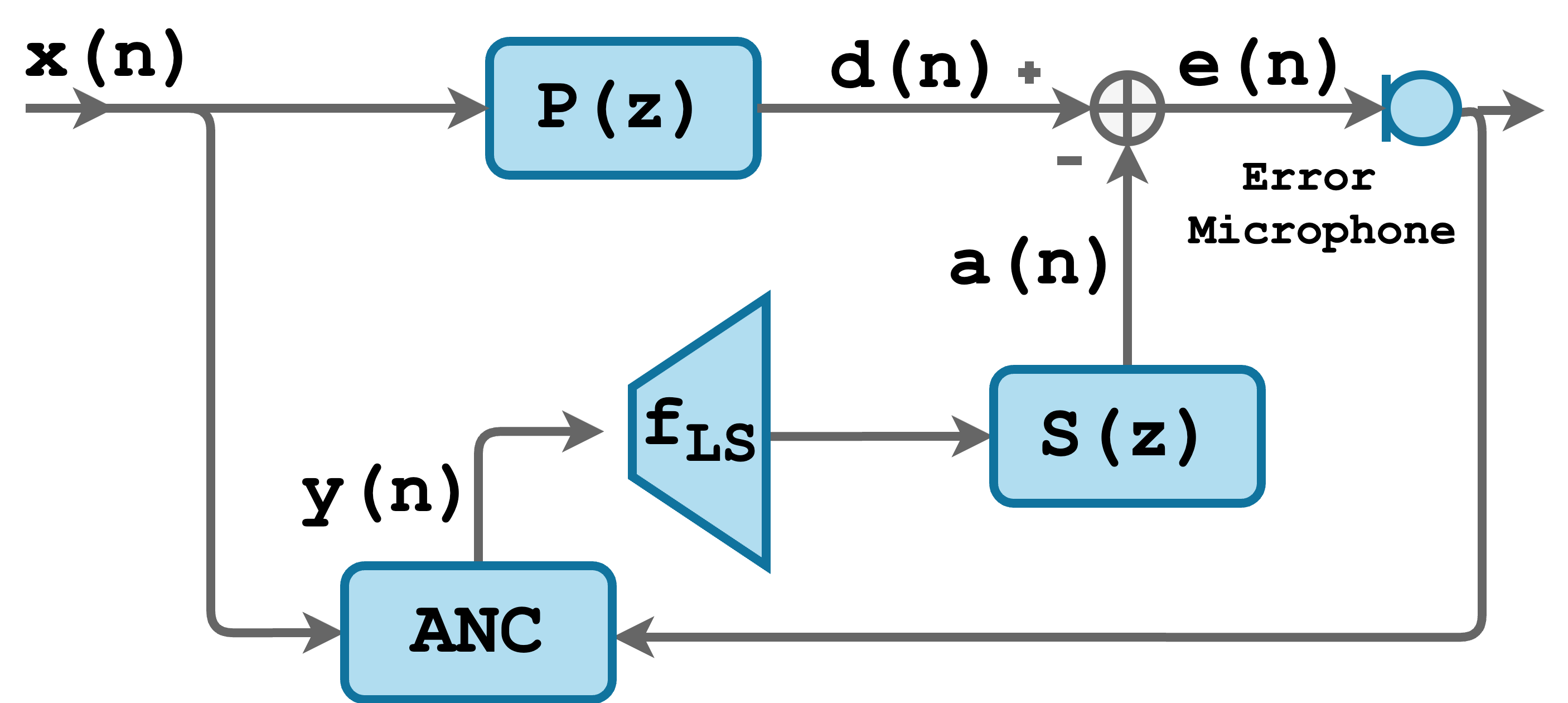}
    \caption{A feedforward ANC setup diagram.}
    \label{figs:anc}
  \end{subfigure}
  \hfill
  \begin{subfigure}[t]{0.49\textwidth}
    \centering
    \includegraphics[width=\textwidth]{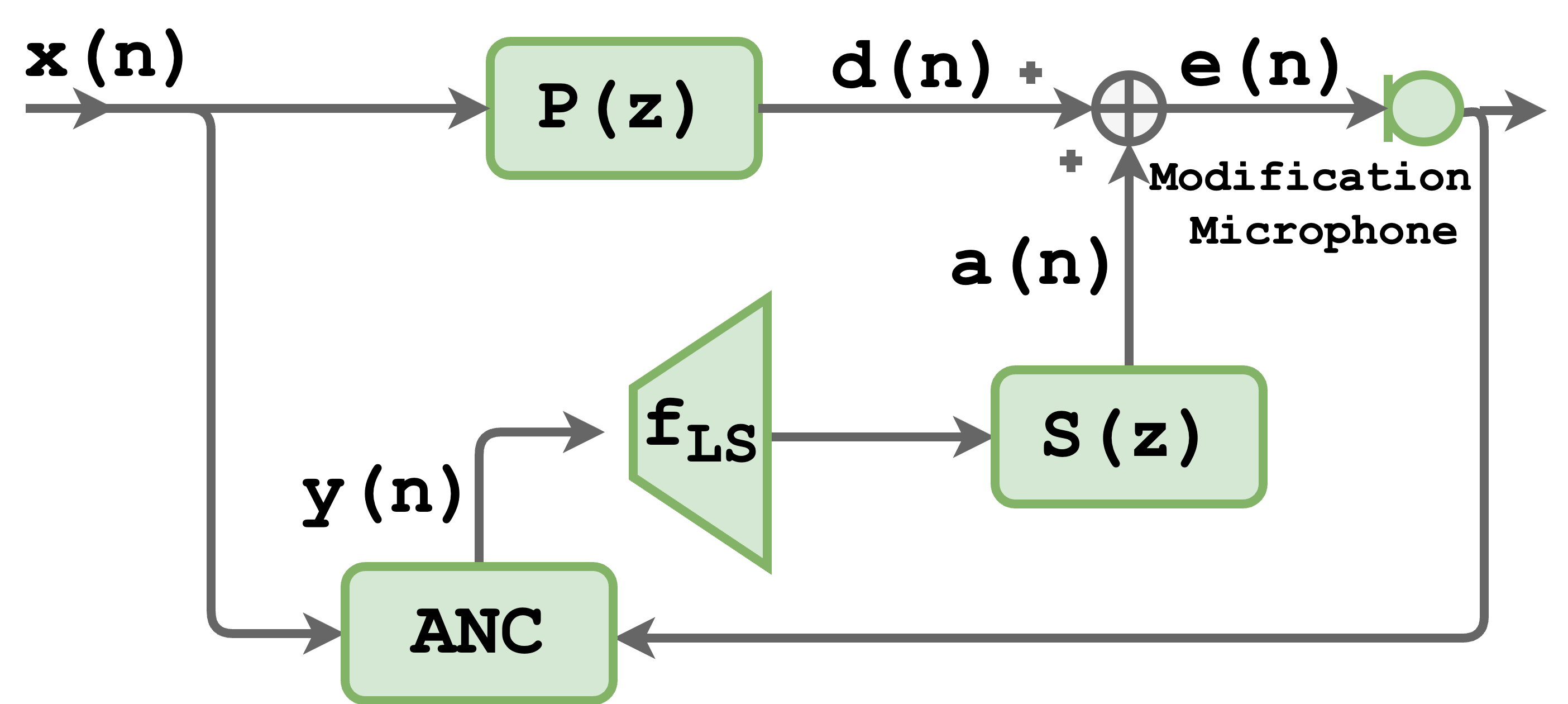}
    \caption{A feedforward ASE setup diagram.}
    \label{figs:ase}
  \end{subfigure}
  \caption{Comparison of feedforward ANC and ASE setups.}
  \label{fig:anc_ase}
\end{figure}

We first examine a feedforward ANC algorithm that employs a single error microphone to lay the foundation for our new ASE framework. In the ANC framework (Figure~\ref{fig:anc_ase}a), the \emph{primary path}, characterized by the transfer function \(P(z)\), models the acoustic propagation from the disturbance source to the error microphone. The \emph{secondary path}, denoted by \(S(z)\), describes the transfer from the loudspeaker to the error microphone. Let \(x(n)\) denote the reference signal applied to the ANC system. The primary signal \(d(n)\) is obtained by filtering \(x(n)\) through the primary path:
\begin{equation}
    d(n) \;=\; P(z)\,*\,x(n)\,,
\end{equation}

where $*$ denotes the convolution operation. The error microphone captures the residual signal \(e(n)\), representing the difference between the original disturbance and the cancellation signal. Both \(x(n)\) and \(e(n)\) are used by the ANC algorithm to compute the canceling signal \(y(n)\). The loudspeaker implements \(y(n)\) according to its electro‐acoustic transfer function \(f_{\mathrm{LS}}\{\cdot\}\). After propagation through the secondary path, the anti-signal (or cancellation signal) is given by
\begin{equation}
    a(n) \;=\; S(z)\,*\,f_{\mathrm{LS}}\bigl(y(n)\bigr)\,.
\label{eq:anti-signal}
\end{equation}

The error signal is defined formally as the difference between the primary signal and the anti-signal:

\begin{equation}
    e(n) \;=\; d(n) - a(n)\,.
\label{eq:eh-signal}
\end{equation}

The objective of the ANC algorithm is to generate \(y(n)\) such that \(e(n)\) is minimized, ideally achieving \(e(n)=0\), which corresponds to complete cancellation of the disturbance. In contrast, the \textbf{ASE framework} uses the primary and anti-signals to construct an enhanced signal:
\begin{equation}
    eh(n) \;=\; d(n) + a(n)\,.
\end{equation}

While ANC aims to eliminate the disturbance, ASE seeks to recover clean speech from a noisy mixture. The objective of the ASE task is to generate \(eh(n)\) such that its deviation from the clean target signal \(c(n)\), i.e., \(eh(n) - c(n)\), is minimized. As illustrated in Figure~\ref{fig:anc_ase}b, the feedforward ASE setup comprises a disturbance source, a reference signal path, and a control filter operating through the secondary path. Given the nature of the task, the error microphone serves as the modification microphone in our framework.

Our work adapts three speech distortion types, previously defined by VoiceFixer~\citep{liu2022voicefixer} for general speech restoration, to the context of our ASE framework. Specifically, our ASE-TM model targets the restoration of speech \(s(n)\) degraded by: \textbf{(i) Additive noise}: This common distortion, where unwanted background sounds obscure the speech, is modeled as the sum of the clean speech signal $s(n)$ and a noise signal $n(n)$:
    \begin{equation}
        d_{\text{noise}}(s(n)) = s(n) + n(n)\,. \label{eq:additive_noise}
    \end{equation}
\textbf{(ii) Reverberation}: Caused by sound reflections in an enclosure, reverberation blurs speech signals. It is modeled by convolving the speech signal $s(n)$ with a room impulse response (RIR) $r(n)$:
    \begin{equation}
        d_{\text{rev}}(s(n)) = s(n) * r(n)\,. \label{eq:reverberation}
    \end{equation}
\textbf{(iii) Clipping}: This distortion arises when signal amplitudes exceed the maximum recordable level, typically due to microphone limitations. Clipping truncates the signal $s(n)$ within a certain range $[-\eta, +\eta]$:
    \begin{equation}
        d_{\text{clip}}(s(n)) = \max(\min(s(n), \eta), -\eta)\,, \quad \eta \in [0,1]\,. \label{eq:clipping}
    \end{equation}
This leads to harmonic distortions and can degrade speech intelligibility. To assess the performance of ASE-TM across these enhancement tasks, we employ a suite of established objective metrics. Consistent with the evaluation protocol in SEmamba~\citep{chao2024investigation}, these include the Wide-band PESQ~\citep{rix2001perceptual}, STOI~\citep{taal2010short}, and the composite measures CSIG (predicting signal distortion), CBAK (predicting background intrusiveness), and COVL (predicting overall speech quality)~\citep{hu2007evaluation}. Furthermore, we incorporate the Normalized Mean Square Error (NMSE), a traditionally well-established metric in the ANC task. The NMSE between a target signal $u(n)$ and an estimated signal $v(n)$ is defined in decibels (dB) as:
\begin{equation}
    \text{NMSE}[u,v] = 10 \cdot \log_{10}\left(\frac{\sum_{n=1}^{M}(u(n)-v(n))^{2}}{\sum_{n=1}^{M}u(n)^{2}}\right)\,, \label{eq:nmse_definition}
\end{equation}
where $M$ is the total number of samples. In our evaluations, $u(n)$ represents the clean target speech $c(n)$, and $v(n)$ is the enhanced speech $eh(n)$ produced by our model (the precise definition of $c(n)$ for each task is detailed in Section~\ref{sec:optimization_objective}).

\section{Method}
\subsection{ASE-TM Architecture Overview}
\begin{figure}[t]
\centering
\includegraphics[width=1\textwidth]{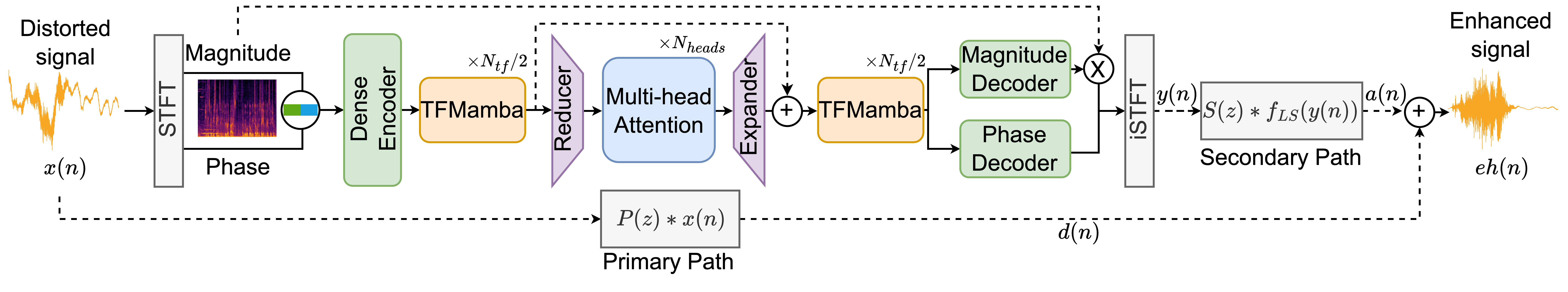}
\caption{ASE-TM Architecture.}
\label{fig:ASE_TM_arc}
\end{figure}

The proposed model, ASE Transformer-Mamba (ASE-TM, Figure~\ref{fig:ASE_TM_arc}), adopts and extends the fundamental structure of the SEmamba architecture~\citep{chao2024investigation}, which consists of a dense encoder, a series of Time-Frequency Mamba (TFMamba) blocks~\citep{xiao2024tf}, and parallel magnitude and phase decoders. A notable distinction of our ASE-TM model is the utilization of Mamba2 blocks~\citep{dao2024transformers} within these TFMamba pathways, in contrast to the original SEmamba architecture, which employed an earlier version of Mamba~\citep{gu2023mamba}. This choice is motivated by the potential improvements in efficiency offered by Mamba2.

The input noisy waveform, \(x(n)\), sampled at a rate of $F_s$, is processed through an STFT. The STFT employs a Hann window of $N_{\text{win}}$ samples, a hop length of $N_{\text{hop}}$ samples, and an $N_{\text{FFT}}$-point FFT, resulting in $N_{\text{freq}} = \floor{N_{\text{win}}/2} + 1$ frequency features per frame. Its magnitude and phase components are horizontally stacked and fed into the network. The dense encoder utilizes convolutional layers and dense blocks to extract initial features from the stacked magnitude and phase, outputting a representation with $C_{\text{enc}}$ channels, each with $N_{\text{enc}}$ features.

The core of the temporal and spectral modeling is based on $N_{\text{tf}}$ TFMamba blocks. Each TFMamba block contains separate Mamba-based pathways (\texttt{time-mamba} and \texttt{freq-mamba}) employing bidirectional Mamba layers to capture dependencies across time and frequency dimensions, respectively~\citep{chao2024investigation, xiao2024tf}.

Following the initial $N_{\text{tf}}/2$ TFMamba blocks, inspired by hybrid approaches like Jamba~\citep{lieber2024jamba}, we introduce an attention-based block. Before applying attention, the feature representation, with $C_{\text{enc}}$ channels, undergoes dimensionality reduction. A 2D convolution with a kernel size of $(1, N_{\text{enc}}/2 + 1)$ reduces the channel dimension from $C_{\text{enc}}$ to $C_{\text{enc}}/4$ and each channel features size from $N_{\text{enc}}$ to $N_{\text{enc}}/2$. This results in a compact representation of size $N_{\text{enc}}/2 \times C_{\text{enc}}/4$ for the attention layer. In addition, positional encoding is applied to this compact representation. A standard Multi-Head Attention layer with $N_{\text{heads}}$ heads is then used on this reduced representation to weigh features based on global context. Following the attention layer, an expansion module employing a transposed 2D convolution, also with a kernel size of $(1, N_{\text{enc}}/2 + 1)$, is used to restore the channel dimension to $C_{\text{enc}}$ and expand the feature dimension back towards $N_{\text{enc}}$ before passing the features to the remaining $N_{\text{tf}}/2$ TFMamba blocks. An additional step before applying the remaining TFMamba blocks is the use of a residual connection that sums the feature representations from before the dimensionality reduction with those after the attention and expansion modules

The magnitude and phase decoders retain the structure used in SEmamba, employing dense blocks and convolutional layers (including transposed convolutions) to reconstruct the target representation—not before applying a residual connection that performs element-wise multiplication between the original STFT magnitude and the predicted magnitude. However, instead of predicting the enhanced spectra directly, the network is trained to output the complex spectrum of the cancelling signal $y(n)$. This signal $y(n)$, after undergoing the electro-acoustic transfer function $f_{\mathrm{LS}}\{\cdot\}$ and propagation through the secondary path $S(z)$, becomes the anti-signal $a(n)$ (as defined in Eq.~\ref{eq:anti-signal}). This anti-signal $a(n)$ is then summed with the primary path signal $d(n)$ to produce the final enhanced signal $eh(n)$ (as defined in Eq.~\ref{eq:eh-signal}).

\label{sec:optimization_objective}
\subsection{Optimization Objective}\label{sec:optimization_objective}
The primary goal of the ASE-TM model is to generate an enhanced signal $eh(n)$ that is as close as possible to a clean target speech signal $c(n)$. The definition of this target $c(n)$ varies based on the specific enhancement task. For \textbf{additive noise reduction}, $c(n)$ is the clean speech signal convolved with the primary path $P(z)$, representing the clean signal as perceived at the modification microphone. For \textbf{dereverberation} and \textbf{declipping}, $c(n)$ is the original anechoic, unclipped clean speech signal, prior to any acoustic path effects or clipping distortion.

The training of ASE-TM largely follows the multi-level loss framework established in SEmamba~\citep{chao2024investigation} and originating from MP-SENet~\citep{lu2023mp}. This framework combines several loss components. Our approach incorporates this established framework with specific modifications and additions. The overall generator loss $\mathcal{L}_{G}$ is a weighted sum of the following components:

\begin{enumerate}
    \item \textbf{Time-Domain Loss ($\mathcal{L}_{\text{Time}}$)}: As an innovation in our work, we employ a combination of L1 and L2 distances between the enhanced waveform $eh(n)$ and the target waveform $c(n)$:
    \begin{equation}
        \mathcal{L}_{\text{Time}} = ||eh(n) - c(n)||_{1} + ||eh(n) - c(n)||_{2}^{2}\,.
    \end{equation}
    This hybrid loss aims to leverage the robustness of L1 to outliers and the smoothness encouraged by L2.

    \item \textbf{Magnitude Spectrum Loss ($\mathcal{L}_{\text{Mag}}$)}: Similar to the time-domain loss, our second innovation involves a combined L1 and L2 loss on the magnitude spectra. If $EN_m$ and $C_m$ are the magnitude spectra of $eh(n)$ and $c(n)$ respectively, then:
    \begin{equation}
        \mathcal{L}_{\text{Mag}} = ||EN_m - C_m||_{1} + ||EN_m - C_m||_{2}^{2}\,.
    \end{equation}
    This contrasts with the L2 loss typically used in MP-SENet for this component.
    \item \textbf{Complex Spectrum Loss ($\mathcal{L}_{\text{Com}}$)}: This loss penalizes differences in the STFT domain. It is the sum of L2 losses on the real and imaginary parts of the STFTs of $eh(n)$ and $c(n)$.
    \item \textbf{Anti-Wrapping Phase Loss ($\mathcal{L}_{\text{Pha}}$)}: This includes instantaneous phase loss, group delay loss, and instantaneous angular frequency loss to directly optimize the phase spectrum, addressing phase wrapping issues.
    \item \textbf{Metric-Based Adversarial Loss ($\mathcal{L}_{\text{Metric}}$)}: This involves a discriminator trained to predict a perceptual metric (e.g., PESQ), guiding the generator to produce outputs that score well on this metric.
    \item \textbf{Consistency Loss ($\mathcal{L}_{\text{Consist}}$)}: We incorporate a consistency loss. This loss minimizes the discrepancy between the complex spectrum directly output by the model's decoders (magnitude and phase) and the complex spectrum obtained by applying STFT to the time-domain waveform $eh(n)$ that results from an inverse STFT of the initially predicted spectrum.
\end{enumerate}
The total generator loss is then defined with the hyperparameter $\gamma$ as follows:
\begin{equation}
\mathcal{L}_{G} = \gamma_{\text{1}}\mathcal{L}_{\text{Time}} + \gamma_{\text{2}}\mathcal{L}_{\text{Mag}} + \gamma_{\text{3}}\mathcal{L}_{\text{Com}} + \gamma_{\text{4}}\mathcal{L}_{\text{Metric}} + \gamma_{\text{5}}\mathcal{L}_{\text{Pha}} + \gamma_{\text{6}}\mathcal{L}_{\text{Consist}}.
\label{eq:loss_total_updated}
\end{equation}

\section{Experiments}

\subsection{Datasets and Task Generation}
\label{sec:datasets}
Our evaluations are conducted across three primary speech restoration tasks: additive noise reduction, dereverberation, and declipping. For the \textbf{additive noise reduction} task, we utilize the VoiceBank-DEMAND dataset~\citep{botinhao2016investigating}, a standard benchmark in speech enhancement. This dataset combines clean speech from the VoiceBank corpus~\citep{veaux2013voice} with various non-stationary noises from the DEMAND database~\citep{thiemann2013diverse}. Our training set consists of utterances from 28 speakers with 10 different noise types at Signal-to-Noise Ratios (SNRs) of 0, 5, 10, and 15~dB. We used two speakers from the training set as the validation set. The test set comprises 824 utterances from 2 unseen speakers, mixed with 5 different unseen noise types at SNRs of 2.5, 7.5, 12.5, and 17.5~dB.

The datasets for the \textbf{dereverberation} and \textbf{declipping} tasks are generated using the clean speech utterances of the speakers available in the VoiceBank corpus. \textbf{Dereverberation}: Reverberant speech is synthesized by convolving the clean VoiceBank utterances with Room Impulse Responses (RIRs) as defined in Eq.~\ref{eq:reverberation} using the SpeechBrain package~\citep{ravanelli2024open}. For training, we randomly sample RIRs from the training portion of the RIR dataset provided alongside VoiceFixer~\citep{liu2022voicefixer}. For the test set, a fixed and distinct set of RIRs (from the VoiceFixer RIR test set) is applied to the clean test utterances to ensure consistent evaluation conditions. \textbf{Declipping}: Clipped speech signals are generated by applying a clipping threshold $\eta$ to the clean utterances according to Eq.~\ref{eq:clipping}. During training, the clipping ratio $\eta$ is uniformly sampled from the range $[0.1, 0.5]$ for each utterance to expose the model to varying degrees of distortion. For testing, a specific clipping threshold is used.

\subsection{Acoustic Path Simulation}
\label{sec:acoustic_simulation}
To emulate the acoustic environment for the ASE framework, we simulate the primary path $P(z)$ and secondary path $S(z)$. Our simulation setup is based on previous setups for the ANC task~\citep{zhang2021deep, zhang2023low}, and it models a rectangular enclosure with dimensions of $3 \times 4 \times 2$ meters (width $\times$ length $\times$ height). Room Impulse Responses (RIRs) are generated using the image method~\citep{allen1979image}, implemented with a Python-based RIR generator package~\citep{habets2006room} with the high-pass filtering option enabled. The modification microphone, capturing $eh(n)$, is simulated at position $[1.5, 3, 1]$ meters. The reference microphone, capturing $x(n)$, is at $[1.5, 1, 1]$ meters, and the cancellation loudspeaker, which outputs the signal leading to $a(n)$, is located at $[1.5, 2.5, 1]$ meters within the enclosure. The length of the simulated RIRs for both $P(z)$ and $S(z)$ is $L_{\text{RIR}} = 512$ taps. The non-linear characteristics of the loudspeaker are modeled using the Scaled Error Function (SEF), defined as $f_{\mathrm{LS}}\{y\} = \int_{0}^{y} \exp(-z^2 / (2\lambda^2)) dz$. Here, $y$ is the input signal to the loudspeaker, and $\lambda^2$ controls the severity of the saturation non-linearity. Different values of $\lambda^2$ are used to simulate varying degrees of distortion, with larger values approaching linear behavior. To introduce variability during training, the room's reverberation time ($T_{60}$) and $\lambda^2$ are randomly sampled from the sets $\{0.15, 0.175, 0.2, 0.225, 0.25\}$ seconds and $\{0.1, 1, 10, \infty \}$, respectivly, for each training sample. For testing, fixed $T_{60}$ and $\lambda^2$ are used.

\subsection{Model Hyperparameters and Training}
\label{sec:implementation_details}
The ASE-TM model processes audio sampled at $F_s=16$~kHz. For the STFT, we use a Hann window of $N_{\text{win}}=400$ samples, a hop length of $N_{\text{hop}}=100$ samples, and an $N_{\text{FFT}}=400$-point FFT. The dense encoder outputs a feature representation with $C_{\text{enc}}=128$ channels, where each channel has a feature dimension of $N_{\text{enc}}=100$. Our model employs a total of $N_{\text{tf}}=8$ TFMamba blocks. The Multi-Head Attention layer within the attention-based block uses $N_{\text{heads}}=10$ heads. Other internal architectural details for the TFMamba blocks and dense convolutional blocks largely follow the configurations presented in SEmamba~\citep{chao2024investigation}. The ASE-TM model is trained for $350$ epochs using the AdamW optimizer~\citep{loshchilov2017decoupled} with $\beta_1 = 0.8$ and $\beta_2 = 0.99$. The initial learning rate is set to $5 \times 10^{-4}$. We use a batch size of $4$. Audio segments of $32,000$ samples (equivalent to 2 seconds at 16~kHz) are used for training. The model parameters yielding the best performance on the validation set, evaluated based on the PESQ score, are saved for final testing. We used an NVIDIA RTX A6000 GPU (internal cluster). The training runtime of the ASE-TM model was $\sim10$ days. 

\subsection{Baseline Methods}
\label{sec:baselines}
We compare our proposed ASE-TM model with several established baseline methods commonly used in ANC. These include THF-FxLMS~\citep{ghasemi2016nonlinear}, which is an extension to the traditional FxLMS algorithm~\citep{kuo1999active}, DeepANC that utilizes a convolutional LSTMs~\citep{zhang2021deep}, and ARN that incorporates an attention mechanism~\citep{zhang2023low}. These baseline methods were adapted and retrained or configured to the ASE framework across the denoising, dereverberation, and declipping tasks.

\section{Results and Analysis}
\label{sec:results}

\subsection{Active Denoising Performance}
The speech denoising performance on the VoiceBank-DEMAND dataset is detailed in Table~\ref{tab:speech_enhancement}. Our ASE-TM model demonstrates superior performance, achieving a PESQ score of $2.98$. This significantly surpasses the baselines: THF-FxLMS achieved a PESQ of $2.37$, and the deep learning-based ANC methods, DeepANC and ARN, yielded PESQ scores of $1.48$ and $2.45$, respectively. These results demonstrate a substantial performance gap between conventional ANC approaches and our ASE-TM, which benefits from actively shaping the speech signal in addition to noise suppression, as also evidenced by its leading scores in CSIG, CBAK, COVL, STOI, and particularly a significantly better NMSE of $-21.76$~dB.

\begin{table}[h]
\centering
\caption{Average denoising results on the VoiceBank-DEMAND test set ($T_{60}= 0.25s$ and $\lambda^2 = \infty$).}
\label{tab:speech_enhancement}
\begin{tabular}{lcccccc}
\toprule
\textbf{Method} & \textbf{PESQ ($\uparrow$)} & \textbf{CSIG ($\uparrow$)} & \textbf{CBAK ($\uparrow$)} & \textbf{COVL ($\uparrow$)} & \textbf{STOI ($\uparrow$)} & \textbf{NMSE ($\downarrow$)} \\
\midrule
Noisy-speech & 1.97 & 3.50 & 2.55 & 2.75 & 0.92 & -8.44\\
THF-FxLMS~\citep{zhang2021deep} & 2.37 & 3.66 & 2.84 & 3.00 & 0.97 & -15.32\\
DeepANC~\citep{zhang2021deep}& 1.48 & 1.99 & 2.19 & 1.69 & 0.93 & -12.80\\
ARN~\citep{zhang2023low}& 2.45 & 3.64 & 3.13 & 3.03 & 0.97 & -20.64\\
ASE-TM (Ours) & \textbf{2.98} & \textbf{4.21} & \textbf{3.49} & \textbf{3.62} & \textbf{0.99} & \textbf{-21.76}\\
\bottomrule
\end{tabular}
\end{table}

\subsection{Dereverberation and Declipping Performance}
The efficacy of ASE-TM was further evaluated on dereverberation and declipping tasks, with results presented in Table~\ref{tab:dereverberation} and Table~\ref{tab:declipping}, respectively. For dereverberation (Table~\ref{tab:dereverberation}), ASE-TM achieved a PESQ score of $2.43$, a considerable improvement from the reverberant speech baseline (PESQ $1.60$). In contrast, the adapted ANC baselines struggled with this task; THF-FxLMS scored a PESQ of $1.43$, while DeepANC and ARN achieved $1.06$ and $1.35$, respectively. This suggests that these methods, even when retrained or configured for the task, were less capable of effectively adjusting their processes to mitigate reverberation in the ASE framework. Similarly, in the declipping task, with a clipping threshold of $\eta=0.25$ (Table~\ref{tab:declipping}), ASE-TM restored the speech to a PESQ score of $3.09$ from an initial clipped speech PESQ of $2.17$. The baseline methods again showed limited effectiveness: THF-FxLMS (PESQ $1.92$), DeepANC (PESQ $1.05$), and ARN (PESQ $1.67$).

These tasks, particularly where the target $c(n)$ is the original clean speech before any primary path effects, highlight the challenge and efficacy of the ASE approach in not just cancelling an interfering signal but actively restoring a desired signal characteristic.

\begin{table}[h]
\centering
\caption{Average dereverberation results on the reverbed test set ($T_{60}= 0.25s$ and $\lambda^2 = \infty$).}
\label{tab:dereverberation}
\begin{tabular}{lcccccc}
\toprule
\textbf{Method} & \textbf{PESQ ($\uparrow$)} & \textbf{CSIG ($\uparrow$)} & \textbf{CBAK ($\uparrow$)} & \textbf{COVL ($\uparrow$)} & \textbf{STOI ($\uparrow$)} & \textbf{NMSE ($\downarrow$)} \\
\midrule
Reverbed-speech & 1.60 & 2.60 & 1.88 & 2.02 & 0.80 & 2.00 \\
THF-FxLMS~\citep{zhang2021deep} & 1.43 & 2.55 & 1.64 & 1.89 & 0.78 & 4.77 \\
DeepANC~\citep{zhang2021deep} & 1.06 & 1.00 & 1.00 & 1.00 & 0.53 & 21.19 \\
ARN~\citep{zhang2023low} & 1.35 & 1.25 & 1.54 & 1.18 & 0.76 & 4.58 \\
ASE-TM (Ours) & \textbf{2.43} & \textbf{3.71} & \textbf{2.67} & \textbf{3.07} & \textbf{0.93} & \textbf{-0.04}\\
\bottomrule
\end{tabular}
\end{table}

\begin{table}[h]
\centering
\caption{Average declipping results on the clipped test set ($\eta = 0.25$, $T_{60}= 0.25s$, and $\lambda^2 = \infty$).}
\label{tab:declipping}
\begin{tabular}{lcccccc}
\toprule
\textbf{Method} & \textbf{PESQ ($\uparrow$)} & \textbf{CSIG ($\uparrow$)} & \textbf{CBAK ($\uparrow$)} & \textbf{COVL ($\uparrow$)} & \textbf{STOI ($\uparrow$)} & \textbf{NMSE ($\downarrow$)} \\
\midrule
Clipped-speech & 2.17 & 3.49 & 2.51 & 2.82 & 0.89 & -0.23 \\
THF-FxLMS~\citep{zhang2021deep} & 1.92 & 3.35 & 2.36 & 2.62 & 0.88 & -0.02 \\
DeepANC~\citep{zhang2021deep} & 1.05 & 1.00 & 1.00 & 1.00 & 0.53 & 11.10 \\
ARN~\citep{zhang2023low} & 1.67 & 1.60 & 2.12 & 1.57 & 0.87 & -0.31 \\
ASE-TM (Ours) & \textbf{3.09} & \textbf{4.20} & \textbf{3.06} & \textbf{3.67} & \textbf{0.93} & \textbf{-1.70} \\
\bottomrule
\end{tabular}
\end{table}

\subsection{Denoising ASE-TM Model Analysis}
\begin{figure}[t]
  \centering
  \begin{subfigure}[t]{0.49\textwidth}
    \centering
    \includegraphics[width=\textwidth]{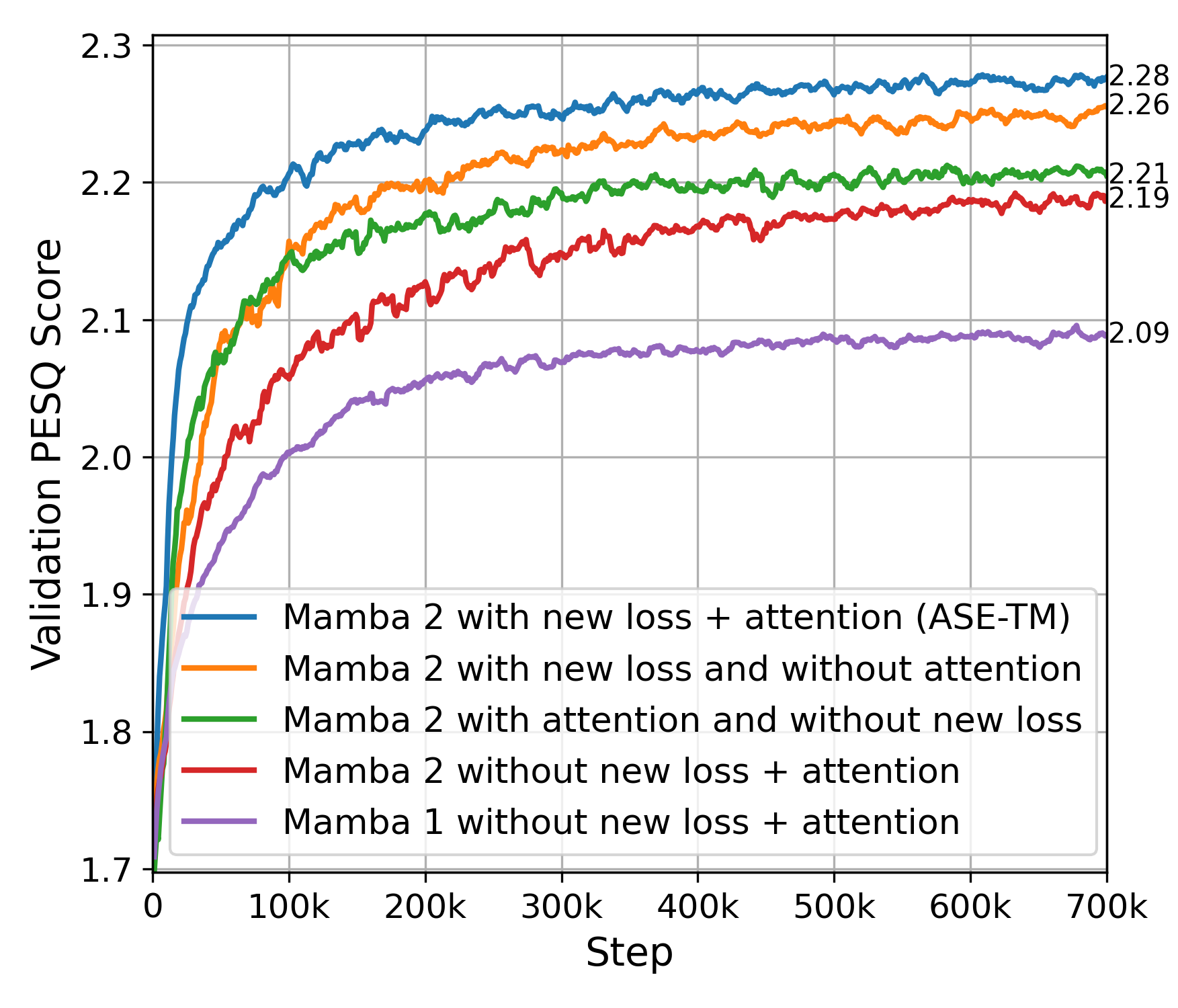}
    \caption{An ablation study.}
    \label{figs:ablation}
  \end{subfigure}
  \begin{subfigure}[t]{0.49\textwidth}
    \centering
    \includegraphics[width=\textwidth]{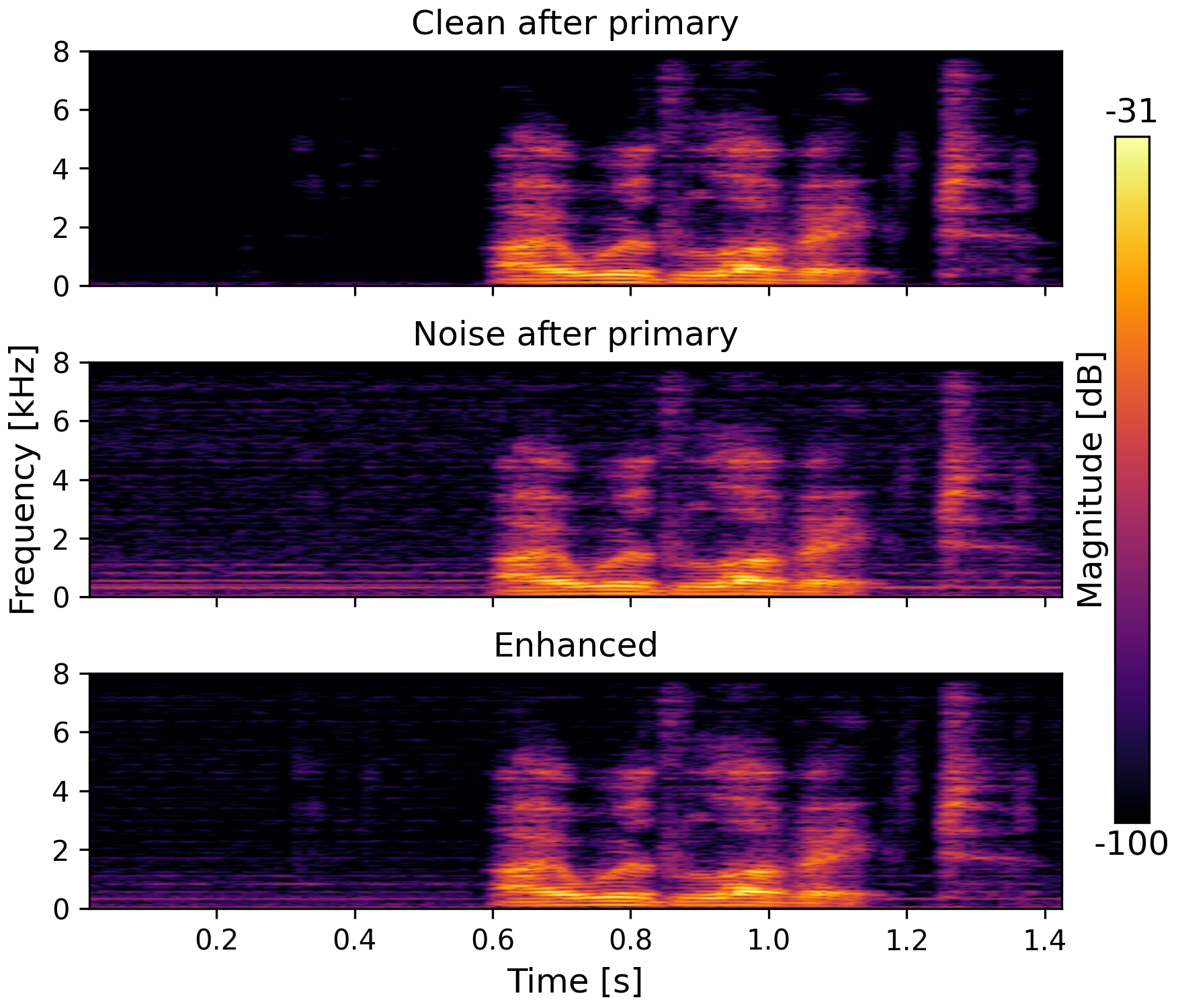}
    \caption{An enhanced (denoised) signal spectrogram.}
    \label{figs:spectrogram}
  \end{subfigure}
  \caption{Model analysis of ASE-TM model for the denoising task. In the ablation study, a moving average with window size $=10$ was applied.}
  \label{fig:noise_model_anlysis}
\end{figure}

An ablation study, presented in Figure~\ref{fig:noise_model_anlysis}a, investigates the contributions of our proposed loss function modifications, the attention mechanism, and the use of Mamba2 over Mamba1 to the ASE-TM model for the denoising task. The full model consistently achieves the highest validation PESQ score throughout training. Replacing Mamba1 with Mamba2 and the modified loss yielded the most considerable performance improvement among all evaluated components. Notably, configurations incorporating the attention mechanism demonstrate a faster convergence to higher performance levels, suggesting that attention aids in efficiently learning relevant features. Spectrogram analysis of a representative denoising example (Figure~\ref{fig:noise_model_anlysis}b) visually confirms the model's effectiveness; the spectrogram of the enhanced signal closely mirrors that of the clean speech (after primary path), indicating successful noise suppression while preserving essential speech characteristics.

To assess robustness, ASE-TM was evaluated under varying acoustic conditions for the denoising task, with results detailed in Table~\ref{tab:ablation_t60_sef}. This analysis focused on ASE-TM due to its significantly better performance over baselines in Table~\ref{tab:speech_enhancement}. The model shows consistent high performance across different $T_{60}$ values under linear loudspeaker conditions ($\lambda^2=\infty$), achieving a PESQ of $3.02$ for $T_{60}=0.15$s and $3.13$ for $T_{60}=0.20$s. When strong non-nonlinearities are introduced (e.g., $\lambda^2=0.1$ at $T_{60}=0.25$s), the PESQ score is $2.74$, still indicating robust performance. As $\lambda^2$ increases (less non-linearity), performance improves, reaching a PESQ of $2.97$ for $\lambda^2=10$ at $T_{60}=0.25$s.

\subsection{Runtime Analysis}
To satisfy real-time constraints in active systems, we evaluated ASE-TM under a future-frame prediction strategy, following prior work~\citep{zhang2021deep,zhang2023low}. In our setup, the causality condition $T_{\text{ASE-TM}} < T_p - T_s$ evaluates to $T_{\text{ASE-TM}} < \frac{2}{343} - \frac{0.5}{343} \approx 0.0043$ seconds, where $T_p$ and $T_s$ denote the acoustic delays of the primary and secondary paths, respectively. To accommodate the model’s inference latency, we predict 500 future frames (0.03125 seconds), remaining within real-time limits of our computational environment. Despite this future context, performance degradation is minimal: on the VoiceBank-DEMAND test set ($T_{60} = 0.25$ s, $\eta^2 = \infty$), ASE-TM achieves a PESQ of 2.96 and STOI of 0.99—closely matching the non-causal configuration.

\begin{table}[h]
\centering
\caption{Average performance of ASE-TM (denoising task) under varying fixed acoustic conditions ($T_{60}$ and loudspeaker non-linearity factor $\lambda^2$) on the VoiceBank-DEMAND test set.}
\label{tab:ablation_t60_sef}
\begin{tabular}{cccccccc} 
\toprule
\textbf{$T_{60}$ (s)} & \textbf{$\lambda^2$} & \textbf{PESQ ($\uparrow$)} & \textbf{CSIG ($\uparrow$)} & \textbf{CBAK ($\uparrow$)} & \textbf{COVL ($\uparrow$)} & \textbf{STOI ($\uparrow$)} & \textbf{NMSE ($\downarrow$)} \\
\midrule
0.25 & 0.1 & 2.74 & 4.01 & 3.29 & 3.39 & 0.98 & -20.21 \\
0.25 & 1.0 & 2.92 & 4.17 & 3.44 & 3.57 & 0.99 & -21.92 \\
0.25 & 10 & 2.97 & 4.21 & 3.48 & 3.62 & 0.99 & -22.37 \\
\midrule 
0.15 & $\infty$ & 3.02 & 4.22 & 3.50 & 3.65 & 0.98 & -22.31 \\
0.20 & $\infty$ & 3.13 & 4.33 & 3.60 & 3.77 & 0.99 & -22.88 \\
\bottomrule
\end{tabular}
\end{table}

\subsection{Dereverberation and Declipping ASE-TM Model Analysis}
\begin{figure}[t]
\centering
\includegraphics[width=0.9\textwidth]{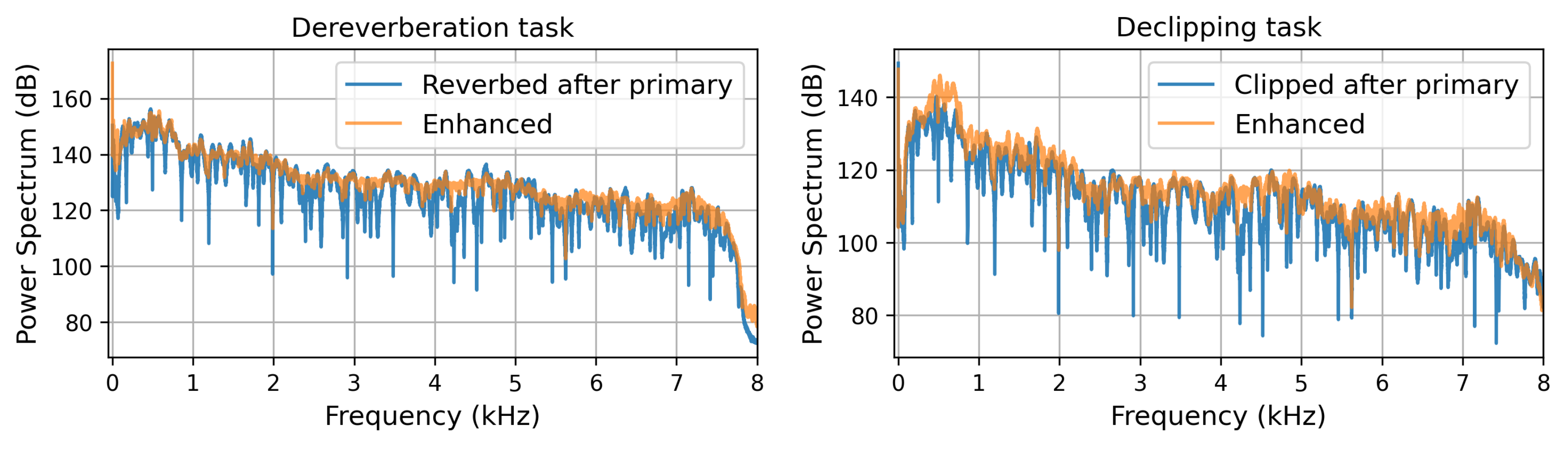}
\caption{Power spectra for the dereverberation and declipping tasks over the entire test set.}
\label{figs:spectrum}
\end{figure}

Figure~\ref{figs:spectrum} presents the power spectra of the enhanced signals for the dereverberation and declipping tasks, over the entire test set. For both tasks, the spectrum of the enhanced signal $eh(n)$ exhibits significantly more power across a broad range of frequencies compared to the distorted input (reverberated or clipped after primary path), indicating successful signal restoration and enrichment. In the declipping task, it is particularly noteworthy that lower frequencies, crucial for speech intelligibility, show substantial power recovery in the enhanced signal's spectrum. We further evaluated the declipping performance under a more aggressive clipping threshold of $\eta=0.1$. The unprocessed clipped speech at this level yielded a PESQ score of $1.53$ and an NMSE of $-0.18$~dB. ASE-TM model restored these signals to a PESQ of $2.52$ (CSIG $3.61$, CBAK $2.76$, COVL $3.08$, STOI $0.91$) and an NMSE of $-1.22$~dB. While these results are lower than for $\eta=0.25$, they represent a substantial improvement over the severely clipped input, showing ASE-TM's capability to handle extreme distortions.

\section{Conclusions and Limitations}
In this paper, we introduced ASE, a novel paradigm that extends beyond traditional ANC by actively shaping the speech signal to enhance quality and intelligibility. Our ASE-TM model, leveraging a Transformer-Mamba architecture and a specialized loss function, demonstrated strong performance across various tasks, including denoising, dereverberation, and declipping, outperforming ANC-centric baseline methods. However, this study also reveals several limitations that require further investigation. The baseline methods, primarily designed for ANC, were adapted to the ASE tasks, which may explain their reduced performance. Furthermore, our current model is trained separately for each task. Future work should explore the development of a unified model capable of handling multiple speech enhancement objectives, potentially leading to more versatile and efficient systems.

If the paradigm of ASE is adopted, it can eventually improve communication for people with hearing impairments and enhance voice communication in devices. However, it could be misused to alter recordings, raising ethical concerns about authenticity and privacy.

\clearpage
\bibliographystyle{unsrtnat} 
\bibliography{neurips_2025}

\begin{thebibliography}{135}
\providecommand{\natexlab}[1]{#1}
\providecommand{\url}[1]{\texttt{#1}}
\expandafter\ifx\csname urlstyle\endcsname\relax
  \providecommand{\doi}[1]{doi: #1}\else
  \providecommand{\doi}{doi: \begingroup \urlstyle{rm}\Url}\fi

\bibitem[Boll(2003)]{boll2003suppression}
Steven Boll.
\newblock Suppression of acoustic noise in speech using spectral subtraction.
\newblock \emph{IEEE Transactions on acoustics, speech, and signal processing}, 27\penalty0 (2):\penalty0 113--120, 2003.

\bibitem[Lim and Oppenheim(1978)]{lim1978all}
Jae Lim and Alan Oppenheim.
\newblock All-pole modeling of degraded speech.
\newblock \emph{IEEE Transactions on Acoustics, Speech, and Signal Processing}, 26\penalty0 (3):\penalty0 197--210, 1978.

\bibitem[Paliwal et~al.(2012)Paliwal, Schwerin, and W{\'o}jcicki]{paliwal2012speech}
Kuldip Paliwal, Belinda Schwerin, and Kamil W{\'o}jcicki.
\newblock Speech enhancement using a minimum mean-square error short-time spectral modulation magnitude estimator.
\newblock \emph{Speech Communication}, 54\penalty0 (2):\penalty0 282--305, 2012.

\bibitem[Pascual et~al.(2017)Pascual, Bonafonte, and Serra]{pascual2017segan}
Santiago Pascual, Antonio Bonafonte, and Joan Serra.
\newblock Segan: Speech enhancement generative adversarial network.
\newblock \emph{arXiv preprint arXiv:1703.09452}, 2017.

\bibitem[Rethage et~al.(2018)Rethage, Pons, and Serra]{rethage2018wavenet}
Dario Rethage, Jordi Pons, and Xavier Serra.
\newblock A wavenet for speech denoising.
\newblock In \emph{2018 IEEE International Conference on Acoustics, Speech and Signal Processing (ICASSP)}, pages 5069--5073. IEEE, 2018.

\bibitem[Pandey and Wang(2018)]{pandey2018new}
Ashutosh Pandey and DeLiang Wang.
\newblock A new framework for supervised speech enhancement in the time domain.
\newblock In \emph{Interspeech}, pages 1136--1140, 2018.

\bibitem[Hu et~al.(2020)Hu, Liu, Lv, Xing, Zhang, Fu, Wu, Zhang, and Xie]{hu2020dccrn}
Yanxin Hu, Yun Liu, Shubo Lv, Mengtao Xing, Shimin Zhang, Yihui Fu, Jian Wu, Bihong Zhang, and Lei Xie.
\newblock Dccrn: Deep complex convolution recurrent network for phase-aware speech enhancement.
\newblock \emph{arXiv preprint arXiv:2008.00264}, 2020.

\bibitem[Fu et~al.(2019)Fu, Liao, Tsao, and Lin]{fu2019metricgan}
Szu-Wei Fu, Chien-Feng Liao, Yu~Tsao, and Shou-De Lin.
\newblock Metricgan: Generative adversarial networks based black-box metric scores optimization for speech enhancement.
\newblock In \emph{International Conference on Machine Learning}, pages 2031--2041. PmLR, 2019.

\bibitem[Fu et~al.(2021)Fu, Yu, Hsieh, Plantinga, Ravanelli, Lu, and Tsao]{Fu_2021}
Szu-Wei Fu, Cheng Yu, Tsun-An Hsieh, Peter Plantinga, Mirco Ravanelli, Xugang Lu, and Yu~Tsao.
\newblock Metricgan+: An improved version of metricgan for speech enhancement.
\newblock In \emph{Interspeech 2021}. ISCA, August 2021.
\newblock \doi{10.21437/interspeech.2021-599}.
\newblock URL \url{http://dx.doi.org/10.21437/interspeech.2021-599}.

\bibitem[Kim et~al.(2021)Kim, Yoon, Cheon, Kang, and Kim]{kim2021multi}
Hyung~Yong Kim, Ji~Won Yoon, Sung~Jun Cheon, Woo~Hyun Kang, and Nam~Soo Kim.
\newblock A multi-resolution approach to gan-based speech enhancement.
\newblock \emph{Applied Sciences}, 11\penalty0 (2):\penalty0 721, 2021.

\bibitem[Shin et~al.(2023)Shin, Lee, Kim, Park, and Han]{shin2023metricgan}
Wooseok Shin, Byung~Hoon Lee, Jin~Sob Kim, Hyun~Joon Park, and Sung~Won Han.
\newblock Metricgan-okd: multi-metric optimization of metricgan via online knowledge distillation for speech enhancement.
\newblock In \emph{International Conference on Machine Learning}, pages 31521--31538. PMLR, 2023.

\bibitem[Shetu et~al.(2025)Shetu, Habets, and Brendel]{shetu2025gan}
Shrishti~Saha Shetu, Emanu{\"e}l~AP Habets, and Andreas Brendel.
\newblock Gan-based speech enhancement for low snr using latent feature conditioning.
\newblock In \emph{ICASSP 2025-2025 IEEE International Conference on Acoustics, Speech and Signal Processing (ICASSP)}, pages 1--5. IEEE, 2025.

\bibitem[Wang et~al.(2021)Wang, He, and Zhu]{Wang_2021}
Kai Wang, Bengbeng He, and Wei-Ping Zhu.
\newblock Tstnn: Two-stage transformer based neural network for speech enhancement in the time domain.
\newblock In \emph{ICASSP 2021 - 2021 IEEE International Conference on Acoustics, Speech and Signal Processing (ICASSP)}, page 7098–7102. IEEE, June 2021.
\newblock \doi{10.1109/icassp39728.2021.9413740}.
\newblock URL \url{http://dx.doi.org/10.1109/ICASSP39728.2021.9413740}.

\bibitem[de~Oliveira et~al.(2022)de~Oliveira, Peer, and Gerkmann]{de_Oliveira_2022}
Danilo de~Oliveira, Tal Peer, and Timo Gerkmann.
\newblock Efficient transformer-based speech enhancement using long frames and stft magnitudes.
\newblock In \emph{Interspeech 2022}, page 2948–2952. ISCA, September 2022.
\newblock \doi{10.21437/interspeech.2022-10781}.
\newblock URL \url{http://dx.doi.org/10.21437/Interspeech.2022-10781}.

\bibitem[Zhang et~al.(2022{\natexlab{a}})Zhang, Chadwick, Ramos, and Bhattacharya]{zhang2022cross}
Shucong Zhang, Malcolm Chadwick, Alberto Gil~CP Ramos, and Sourav Bhattacharya.
\newblock Cross-attention is all you need: Real-time streaming transformers for personalised speech enhancement.
\newblock \emph{arXiv preprint arXiv:2211.04346}, 2022{\natexlab{a}}.

\bibitem[Cao et~al.(2022)Cao, Abdulatif, and Yang]{cao2022cmgan}
Ruizhe Cao, Sherif Abdulatif, and Bin Yang.
\newblock Cmgan: Conformer-based metric gan for speech enhancement.
\newblock \emph{arXiv preprint arXiv:2203.15149}, 2022.

\bibitem[Ye and Wan(2023)]{ye2023improved}
Moujia Ye and Hongjie Wan.
\newblock Improved transformer-based dual-path network with amplitude and complex domain feature fusion for speech enhancement.
\newblock \emph{Entropy}, 25\penalty0 (2):\penalty0 228, 2023.

\bibitem[Zhang et~al.(2024{\natexlab{a}})Zhang, Zhu, Qian, Ambikairajah, and Li]{Zhang_2024}
Qiquan Zhang, Hongxu Zhu, Xinyuan Qian, Eliathamby Ambikairajah, and Haizhou Li.
\newblock An exploration of length generalization in transformer-based speech enhancement.
\newblock In \emph{Interspeech 2024}, page 1725–1729. ISCA, September 2024{\natexlab{a}}.
\newblock \doi{10.21437/interspeech.2024-1831}.
\newblock URL \url{http://dx.doi.org/10.21437/interspeech.2024-1831}.

\bibitem[Guimarães et~al.(2025)Guimarães, Su, Kumar, Falk, and Jin]{guimares2025ditse}
Heitor~R. Guimarães, Jiaqi Su, Rithesh Kumar, Tiago~H. Falk, and Zeyu Jin.
\newblock Ditse: High-fidelity generative speech enhancement via latent diffusion transformers, 2025.

\bibitem[Lu et~al.(2022)Lu, Wang, Watanabe, Richard, Yu, and Tsao]{lu2022conditional}
Yen-Ju Lu, Zhong-Qiu Wang, Shinji Watanabe, Alexander Richard, Cheng Yu, and Yu~Tsao.
\newblock Conditional diffusion probabilistic model for speech enhancement.
\newblock In \emph{ICASSP 2022-2022 IEEE International Conference on Acoustics, Speech and Signal Processing (ICASSP)}, pages 7402--7406. Ieee, 2022.

\bibitem[Welker et~al.(2022)Welker, Richter, and Gerkmann]{welker2022speech}
Simon Welker, Julius Richter, and Timo Gerkmann.
\newblock Speech enhancement with score-based generative models in the complex stft domain.
\newblock \emph{arXiv preprint arXiv:2203.17004}, 2022.

\bibitem[Richter et~al.(2023)Richter, Welker, Lemercier, Lay, and Gerkmann]{Richter_2023}
Julius Richter, Simon Welker, Jean-Marie Lemercier, Bunlong Lay, and Timo Gerkmann.
\newblock Speech enhancement and dereverberation with diffusion-based generative models.
\newblock \emph{IEEE/ACM Transactions on Audio, Speech, and Language Processing}, 31:\penalty0 2351–2364, 2023.
\newblock ISSN 2329-9304.
\newblock \doi{10.1109/taslp.2023.3285241}.
\newblock URL \url{http://dx.doi.org/10.1109/TASLP.2023.3285241}.

\bibitem[Lemercier et~al.(2023)Lemercier, Richter, Welker, and Gerkmann]{Lemercier_2023}
Jean-Marie Lemercier, Julius Richter, Simon Welker, and Timo Gerkmann.
\newblock Storm: A diffusion-based stochastic regeneration model for speech enhancement and dereverberation.
\newblock \emph{IEEE/ACM Transactions on Audio, Speech, and Language Processing}, 31:\penalty0 2724–2737, 2023.
\newblock ISSN 2329-9304.
\newblock \doi{10.1109/taslp.2023.3294692}.
\newblock URL \url{http://dx.doi.org/10.1109/TASLP.2023.3294692}.

\bibitem[Tai et~al.(2023)Tai, Zhou, Trajcevski, and Zhong]{tai2023revisiting}
Wenxin Tai, Fan Zhou, Goce Trajcevski, and Ting Zhong.
\newblock Revisiting denoising diffusion probabilistic models for speech enhancement: Condition collapse, efficiency and refinement.
\newblock In \emph{Proceedings of the AAAI conference on artificial intelligence}, volume~37, pages 13627--13635, 2023.

\bibitem[Ayilo et~al.(2024)Ayilo, Sadeghi, and Serizel]{Ayilo_2024}
Jean-Eudes Ayilo, Mostafa Sadeghi, and Romain Serizel.
\newblock Diffusion-based speech enhancement with a weighted generative-supervised learning loss.
\newblock In \emph{ICASSP 2024 - 2024 IEEE International Conference on Acoustics, Speech and Signal Processing (ICASSP)}, page 12506–12510. IEEE, April 2024.
\newblock \doi{10.1109/icassp48485.2024.10446805}.
\newblock URL \url{http://dx.doi.org/10.1109/ICASSP48485.2024.10446805}.

\bibitem[Lueg(1936)]{lueg1936process}
Paul Lueg.
\newblock Process of silencing sound oscillations.
\newblock \emph{US patent 2043416}, 1936.

\bibitem[Nelson and Elliott(1991)]{nelson1991active}
Philip~Arthur Nelson and Stephen~J Elliott.
\newblock \emph{Active control of sound}.
\newblock Academic press, 1991.

\bibitem[Fuller et~al.(1996)Fuller, Elliott, and Nelson]{fuller1996active}
Christopher~C Fuller, Sharon Elliott, and Philip~Arthur Nelson.
\newblock \emph{Active control of vibration}.
\newblock Academic press, 1996.

\bibitem[Hansen et~al.(1997)Hansen, Snyder, Qiu, Brooks, and Moreau]{hansen1997active}
Colin~H Hansen, Scott~D Snyder, Xiaojun Qiu, Laura~A Brooks, and Danielle~J Moreau.
\newblock \emph{Active control of noise and vibration}.
\newblock E \& Fn Spon London, 1997.

\bibitem[Kuo and Morgan(1999{\natexlab{a}})]{kuo1999active}
Sen~M Kuo and Dennis~R Morgan.
\newblock Active noise control: a tutorial review.
\newblock \emph{Proceedings of the IEEE}, 87\penalty0 (6):\penalty0 943--973, 1999{\natexlab{a}}.

\bibitem[Zhang and Wang(2021)]{zhang2021deep}
Hao Zhang and DeLiang Wang.
\newblock Deep anc: A deep learning approach to active noise control.
\newblock \emph{Neural Networks}, 141:\penalty0 1--10, 2021.

\bibitem[Park et~al.(2023)Park, Choi, Kim, and Chang]{park2023had}
JungPhil Park, Jeong-Hwan Choi, Yungyeo Kim, and Joon-Hyuk Chang.
\newblock Had-anc: A hybrid system comprising an adaptive filter and deep neural networks for active noise control.
\newblock In \emph{Proceedings of the Annual Conference of the International Speech Communication Association, INTERSPEECH}, volume 2023, pages 2513--2517. International Speech Communication Association, 2023.

\bibitem[Mostafavi and Cha(2023)]{mostafavi2023deep}
Alireza Mostafavi and Young-Jin Cha.
\newblock Deep learning-based active noise control on construction sites.
\newblock \emph{Automation in Construction}, 151:\penalty0 104885, 2023.

\bibitem[Cha et~al.(2023)Cha, Mostafavi, and Benipal]{cha2023dnoisenet}
Young-Jin Cha, Alireza Mostafavi, and Sukhpreet~S Benipal.
\newblock Dnoisenet: Deep learning-based feedback active noise control in various noisy environments.
\newblock \emph{Engineering Applications of Artificial Intelligence}, 121:\penalty0 105971, 2023.

\bibitem[Singh et~al.(2024)Singh, Gupta, Kumar, and Bahl]{singh2024enhancing}
Deepali Singh, Rinki Gupta, Arun Kumar, and Rajendar Bahl.
\newblock Enhancing active noise control through stacked autoencoders: Training strategies, comparative analysis, and evaluation with practical setup.
\newblock \emph{Engineering Applications of Artificial Intelligence}, 135:\penalty0 108811, 2024.

\bibitem[Pike and Cheer(2023)]{pike2023generalized}
Alexander Pike and Jordan Cheer.
\newblock Generalized performance of neural network controllers for feedforward active control of nonlinear systems.
\newblock 2023.

\bibitem[Mishaly et~al.(2025)Mishaly, Wolf, and Nachmani]{mishaly2025deepasc}
Yehuda Mishaly, Lior Wolf, and Eliya Nachmani.
\newblock Deep active speech cancellation with multi-band mamba network, 2025.
\newblock URL \url{https://arxiv.org/abs/2502.01185}.

\bibitem[Rix et~al.(2001)Rix, Beerends, Hollier, and Hekstra]{rix2001perceptual}
Antony~W Rix, John~G Beerends, Michael~P Hollier, and Andries~P Hekstra.
\newblock Perceptual evaluation of speech quality (pesq)-a new method for speech quality assessment of telephone networks and codecs.
\newblock In \emph{2001 IEEE international conference on acoustics, speech, and signal processing. Proceedings (Cat. No. 01CH37221)}, volume~2, pages 749--752. IEEE, 2001.

\bibitem[Defossez et~al.(2020)Defossez, Synnaeve, and Adi]{defossez2020real}
Alexandre Defossez, Gabriel Synnaeve, and Yossi Adi.
\newblock Real time speech enhancement in the waveform domain.
\newblock \emph{arXiv preprint arXiv:2006.12847}, 2020.

\bibitem[Gulati et~al.(2020)Gulati, Qin, Chiu, Parmar, Zhang, Yu, Han, Wang, Zhang, Wu, and Pang]{gulati2020conformerconvolutionaugmentedtransformerspeech}
Anmol Gulati, James Qin, Chung-Cheng Chiu, Niki Parmar, Yu~Zhang, Jiahui Yu, Wei Han, Shibo Wang, Zhengdong Zhang, Yonghui Wu, and Ruoming Pang.
\newblock Conformer: Convolution-augmented transformer for speech recognition, 2020.
\newblock URL \url{https://arxiv.org/abs/2005.08100}.

\bibitem[Subakan et~al.(2021{\natexlab{a}})Subakan, Ravanelli, Cornell, Bronzi, and Zhong]{subakan2021attentionneedspeechseparation}
Cem Subakan, Mirco Ravanelli, Samuele Cornell, Mirko Bronzi, and Jianyuan Zhong.
\newblock Attention is all you need in speech separation, 2021{\natexlab{a}}.
\newblock URL \url{https://arxiv.org/abs/2010.13154}.

\bibitem[Zhang et~al.(2022{\natexlab{b}})Zhang, Pandey, and Wang]{zhang2022attentive}
Hao Zhang, Ashutosh Pandey, and DeLiang Wang.
\newblock Attentive recurrent network for low-latency active noise control.
\newblock In \emph{INTERSPEECH}, pages 956--960, 2022{\natexlab{b}}.

\bibitem[Shi et~al.(2022{\natexlab{a}})Shi, Lam, Ooi, Shen, and Gan]{shi2022selective}
Dongyuan Shi, Bhan Lam, Kenneth Ooi, Xiaoyi Shen, and Woon-Seng Gan.
\newblock Selective fixed-filter active noise control based on convolutional neural network.
\newblock \emph{Signal Processing}, 190:\penalty0 108317, 2022{\natexlab{a}}.

\bibitem[Luo et~al.(2022)Luo, Shi, and Gan]{luo2022hybrid}
Zhengding Luo, Dongyuan Shi, and Woon-Seng Gan.
\newblock A hybrid sfanc-fxnlms algorithm for active noise control based on deep learning.
\newblock \emph{IEEE Signal Processing Letters}, 29:\penalty0 1102--1106, 2022.

\bibitem[Ren and Zhang(2022)]{ren2022improved}
Xing Ren and Hongwei Zhang.
\newblock An improved artificial bee colony algorithm for model-free active noise control: algorithm and implementation.
\newblock \emph{IEEE Transactions on Instrumentation and Measurement}, 71:\penalty0 1--11, 2022.

\bibitem[Zhou et~al.(2023)Zhou, Zhao, and Liu]{zhou2023genetic}
Yang Zhou, Haiquan Zhao, and Dongxu Liu.
\newblock Genetic algorithm-based adaptive active noise control without secondary path identification.
\newblock \emph{IEEE Transactions on Instrumentation and Measurement}, 2023.

\bibitem[Luo et~al.(2024{\natexlab{a}})Luo, Shi, Shen, and Gan]{luo2024unsupervised}
Zhengding Luo, Dongyuan Shi, Xiaoyi Shen, and Woon-Seng Gan.
\newblock Unsupervised learning based end-to-end delayless generative fixed-filter active noise control.
\newblock In \emph{ICASSP 2024-2024 IEEE International Conference on Acoustics, Speech and Signal Processing (ICASSP)}, pages 441--445. IEEE, 2024{\natexlab{a}}.

\bibitem[Shi et~al.(2024)Shi, Gan, Shen, Luo, and Ji]{shi2024behind}
Dongyuan Shi, Woon-seng Gan, Xiaoyi Shen, Zhengding Luo, and Junwei Ji.
\newblock What is behind the meta-learning initialization of adaptive filter?—a naive method for accelerating convergence of adaptive multichannel active noise control.
\newblock \emph{Neural Networks}, 172:\penalty0 106145, 2024.

\bibitem[Zhang and Wang(2023)]{zhang2023deep}
Hao Zhang and DeLiang Wang.
\newblock Deep mcanc: A deep learning approach to multi-channel active noise control.
\newblock \emph{Neural Networks}, 158:\penalty0 318--327, 2023.

\bibitem[Anto{\~n}anzas et~al.(2023)Anto{\~n}anzas, Ferrer, De~Diego, and Gonzalez]{antonanzas2023remote}
Christian Anto{\~n}anzas, Miguel Ferrer, Maria De~Diego, and Alberto Gonzalez.
\newblock Remote microphone technique for active noise control over distributed networks.
\newblock \emph{IEEE/ACM Transactions on Audio, Speech, and Language Processing}, 31:\penalty0 1522--1535, 2023.

\bibitem[Xiao et~al.(2023)Xiao, Xu, and Zhao]{xiao2023spatially}
Tong Xiao, Buye Xu, and Chuming Zhao.
\newblock Spatially selective active noise control systems.
\newblock \emph{The Journal of the Acoustical Society of America}, 153\penalty0 (5):\penalty0 2733--2733, 2023.

\bibitem[Zhang et~al.(2023{\natexlab{a}})Zhang, Zhang, Ma, Samarasinghe, and Sun]{zhang2023time}
Huawei Zhang, Jihui Zhang, Fei Ma, Prasanga~N Samarasinghe, and Huiyuan Sun.
\newblock A time-domain multi-channel directional active noise control system.
\newblock In \emph{2023 31st European Signal Processing Conference (EUSIPCO)}, pages 376--380. IEEE, 2023{\natexlab{a}}.

\bibitem[Shi et~al.(2023{\natexlab{a}})Shi, Lam, Shen, and Gan]{shi2023multichannel}
Dongyuan Shi, Bhan Lam, Xiaoyi Shen, and Woon-Seng Gan.
\newblock Multichannel two-gradient direction filtered reference least mean square algorithm for output-constrained multichannel active noise control.
\newblock \emph{Signal Processing}, 207:\penalty0 108938, 2023{\natexlab{a}}.

\bibitem[Zhang et~al.(2023{\natexlab{b}})Zhang, Pandey, et~al.]{zhang2023low}
Hao Zhang, Ashutosh Pandey, et~al.
\newblock Low-latency active noise control using attentive recurrent network.
\newblock \emph{IEEE/ACM transactions on audio, speech, and language processing}, 31:\penalty0 1114--1123, 2023{\natexlab{b}}.

\bibitem[Liu et~al.(2022)Liu, Liu, Kong, Tian, Zhao, Wang, Huang, and Wang]{liu2022voicefixer}
Haohe Liu, Xubo Liu, Qiuqiang Kong, Qiao Tian, Yan Zhao, DeLiang Wang, Chuanzeng Huang, and Yuxuan Wang.
\newblock Voicefixer: A unified framework for high-fidelity speech restoration.
\newblock \emph{arXiv preprint arXiv:2204.05841}, 2022.

\bibitem[Chao et~al.(2024)Chao, Cheng, La~Quatra, Siniscalchi, Yang, Fu, and Tsao]{chao2024investigation}
Rong Chao, Wen-Huang Cheng, Moreno La~Quatra, Sabato~Marco Siniscalchi, Chao-Han~Huck Yang, Szu-Wei Fu, and Yu~Tsao.
\newblock An investigation of incorporating mamba for speech enhancement.
\newblock \emph{arXiv preprint arXiv:2405.06573}, 2024.

\bibitem[Taal et~al.(2010)Taal, Hendriks, Heusdens, and Jensen]{taal2010short}
Cees~H Taal, Richard~C Hendriks, Richard Heusdens, and Jesper Jensen.
\newblock A short-time objective intelligibility measure for time-frequency weighted noisy speech.
\newblock In \emph{2010 IEEE international conference on acoustics, speech and signal processing}, pages 4214--4217. IEEE, 2010.

\bibitem[Hu and Loizou(2007)]{hu2007evaluation}
Yi~Hu and Philipos~C Loizou.
\newblock Evaluation of objective quality measures for speech enhancement.
\newblock \emph{IEEE Transactions on audio, speech, and language processing}, 16\penalty0 (1):\penalty0 229--238, 2007.

\bibitem[Xiao and Das(2024)]{xiao2024tf}
Yang Xiao and Rohan~Kumar Das.
\newblock Tf-mamba: A time-frequency network for sound source localization.
\newblock \emph{arXiv preprint arXiv:2409.05034}, 2024.

\bibitem[Dao and Gu(2024)]{dao2024transformers}
Tri Dao and Albert Gu.
\newblock Transformers are ssms: Generalized models and efficient algorithms through structured state space duality.
\newblock \emph{arXiv preprint arXiv:2405.21060}, 2024.

\bibitem[Gu and Dao(2023)]{gu2023mamba}
Albert Gu and Tri Dao.
\newblock Mamba: Linear-time sequence modeling with selective state spaces.
\newblock \emph{arXiv preprint arXiv:2312.00752}, 2023.

\bibitem[Lieber et~al.(2024)Lieber, Lenz, Bata, Cohen, Osin, Dalmedigos, Safahi, Meirom, Belinkov, Shalev-Shwartz, et~al.]{lieber2024jamba}
Opher Lieber, Barak Lenz, Hofit Bata, Gal Cohen, Jhonathan Osin, Itay Dalmedigos, Erez Safahi, Shaked Meirom, Yonatan Belinkov, Shai Shalev-Shwartz, et~al.
\newblock Jamba: A hybrid transformer-mamba language model.
\newblock \emph{arXiv preprint arXiv:2403.19887}, 2024.

\bibitem[Lu et~al.(2023)Lu, Ai, and Ling]{lu2023mp}
Ye-Xin Lu, Yang Ai, and Zhen-Hua Ling.
\newblock Mp-senet: A speech enhancement model with parallel denoising of magnitude and phase spectra.
\newblock \emph{arXiv preprint arXiv:2305.13686}, 2023.

\bibitem[Botinhao et~al.(2016)Botinhao, Wang, Takaki, and Yamagishi]{botinhao2016investigating}
Cassia~Valentini Botinhao, Xin Wang, Shinji Takaki, and Junichi Yamagishi.
\newblock Investigating rnn-based speech enhancement methods for noise-robust text-to-speech.
\newblock In \emph{9th ISCA speech synthesis workshop}, pages 159--165, 2016.

\bibitem[Veaux et~al.(2013)Veaux, Yamagishi, and King]{veaux2013voice}
Christophe Veaux, Junichi Yamagishi, and Simon King.
\newblock The voice bank corpus: Design, collection and data analysis of a large regional accent speech database.
\newblock In \emph{2013 international conference oriental COCOSDA held jointly with 2013 conference on Asian spoken language research and evaluation (O-COCOSDA/CASLRE)}, pages 1--4. IEEE, 2013.

\bibitem[Thiemann et~al.(2013)Thiemann, Ito, and Vincent]{thiemann2013diverse}
Joachim Thiemann, Nobutaka Ito, and Emmanuel Vincent.
\newblock The diverse environments multi-channel acoustic noise database (demand): A database of multichannel environmental noise recordings.
\newblock In \emph{Proceedings of Meetings on Acoustics}, volume~19. AIP Publishing, 2013.

\bibitem[Ravanelli et~al.(2024)Ravanelli, Parcollet, Moumen, de~Langen, Subakan, Plantinga, Wang, Mousavi, Della~Libera, Ploujnikov, et~al.]{ravanelli2024open}
Mirco Ravanelli, Titouan Parcollet, Adel Moumen, Sylvain de~Langen, Cem Subakan, Peter Plantinga, Yingzhi Wang, Pooneh Mousavi, Luca Della~Libera, Artem Ploujnikov, et~al.
\newblock Open-source conversational ai with speechbrain 1.0.
\newblock \emph{Journal of Machine Learning Research}, 25\penalty0 (333):\penalty0 1--11, 2024.

\bibitem[Allen and Berkley(1979)]{allen1979image}
Jont~B Allen and David~A Berkley.
\newblock Image method for efficiently simulating small-room acoustics.
\newblock \emph{The Journal of the Acoustical Society of America}, 65\penalty0 (4):\penalty0 943--950, 1979.

\bibitem[Habets(2006)]{habets2006room}
Emanuel~AP Habets.
\newblock Room impulse response generator.
\newblock \emph{Technische Universiteit Eindhoven, Tech. Rep}, 2\penalty0 (2.4):\penalty0 1, 2006.

\bibitem[Loshchilov and Hutter(2017)]{loshchilov2017decoupled}
Ilya Loshchilov and Frank Hutter.
\newblock Decoupled weight decay regularization.
\newblock \emph{arXiv preprint arXiv:1711.05101}, 2017.

\bibitem[Ghasemi et~al.(2016)Ghasemi, Kamil, and Marhaban]{ghasemi2016nonlinear}
Sepehr Ghasemi, Raja Kamil, and Mohammad~Hamiruce Marhaban.
\newblock Nonlinear thf-fxlms algorithm for active noise control with loudspeaker nonlinearity.
\newblock \emph{Asian Journal of Control}, 18\penalty0 (2):\penalty0 502--513, 2016.

\bibitem[Bengio and LeCun(2007)]{Bengio+chapter2007}
Yoshua Bengio and Yann LeCun.
\newblock Scaling learning algorithms towards {AI}.
\newblock In \emph{Large Scale Kernel Machines}. MIT Press, 2007.

\bibitem[Hinton et~al.(2006)Hinton, Osindero, and Teh]{Hinton06}
Geoffrey~E. Hinton, Simon Osindero, and Yee~Whye Teh.
\newblock A fast learning algorithm for deep belief nets.
\newblock \emph{Neural Computation}, 18:\penalty0 1527--1554, 2006.

\bibitem[Goodfellow et~al.(2016)Goodfellow, Bengio, Courville, and Bengio]{goodfellow2016deep}
Ian Goodfellow, Yoshua Bengio, Aaron Courville, and Yoshua Bengio.
\newblock \emph{Deep learning}, volume~1.
\newblock MIT Press, 2016.

\bibitem[Tropp(2006)]{tropp2006just}
Joel~A Tropp.
\newblock Just relax: Convex programming methods for identifying sparse signals in noise.
\newblock \emph{IEEE transactions on information theory}, 52\penalty0 (3):\penalty0 1030--1051, 2006.

\bibitem[Iotov et~al.(2022)Iotov, N{\o}rholm, Belyi, Dyrholm, and Christensen]{iotov2022computationally}
Yurii Iotov, Sidsel~Marie N{\o}rholm, Valiantsin Belyi, Mads Dyrholm, and Mads~Gr{\ae}sb{\o}ll Christensen.
\newblock Computationally efficient fixed-filter anc for speech based on long-term prediction for headphone applications.
\newblock In \emph{ICASSP 2022-2022 IEEE International Conference on Acoustics, Speech and Signal Processing (ICASSP)}, pages 761--765. IEEE, 2022.

\bibitem[Iotov et~al.(2023)Iotov, N{\o}rholm, Belyi, and Christensen]{iotov2023adaptive}
Yurii Iotov, Sidsel~Marie N{\o}rholm, Valiantsin Belyi, and Mads~Gr{\ae}sb{\o}ll Christensen.
\newblock Adaptive sparse linear prediction in fixed-filter anc headphone applications for multi-speaker speech reduction.
\newblock In \emph{2023 IEEE Workshop on Applications of Signal Processing to Audio and Acoustics (WASPAA)}, pages 1--5. IEEE, 2023.

\bibitem[Li and Chen(2024)]{li2024spmamba}
Kai Li and Guo Chen.
\newblock Spmamba: State-space model is all you need in speech separation.
\newblock \emph{arXiv preprint arXiv:2404.02063}, 2024.

\bibitem[Zhang et~al.(2024{\natexlab{b}})Zhang, Zhang, Liu, Xiao, Qian, Ahmed, Ambikairajah, Li, and Epps]{zhang2024mamba}
Xiangyu Zhang, Qiquan Zhang, Hexin Liu, Tianyi Xiao, Xinyuan Qian, Beena Ahmed, Eliathamby Ambikairajah, Haizhou Li, and Julien Epps.
\newblock Mamba in speech: Towards an alternative to self-attention.
\newblock \emph{arXiv preprint arXiv:2405.12609}, 2024{\natexlab{b}}.

\bibitem[Jiang et~al.(2024{\natexlab{a}})Jiang, Han, and Mesgarani]{jiang2024dual}
Xilin Jiang, Cong Han, and Nima Mesgarani.
\newblock Dual-path mamba: Short and long-term bidirectional selective structured state space models for speech separation.
\newblock \emph{arXiv preprint arXiv:2403.18257}, 2024{\natexlab{a}}.

\bibitem[Lee and Kim(2024)]{waveumamba}
Yongjoon Lee and Chanwoo Kim.
\newblock Wave-u-mamba: An end-to-end framework for high-quality and efficient speech super resolution.
\newblock \emph{arXiv preprint arXiv:2403.09337}, 2024.

\bibitem[Quan and Li(2024)]{quan2024multichannel}
Changsheng Quan and Xiaofei Li.
\newblock Multichannel long-term streaming neural speech enhancement for static and moving speakers.
\newblock \emph{arXiv preprint arXiv:2403.07675}, 2024.

\bibitem[Erol et~al.(2024)Erol, Senocak, Feng, and Chung]{erol2024audio}
Mehmet~Hamza Erol, Arda Senocak, Jiu Feng, and Joon~Son Chung.
\newblock Audio mamba: Bidirectional state space model for audio representation learning.
\newblock \emph{arXiv preprint arXiv:2406.03344}, 2024.

\bibitem[Mu et~al.(2024)Mu, Zhang, Yue, Wang, Tang, and Yin]{mu2024seld}
Da~Mu, Zhicheng Zhang, Haobo Yue, Zehao Wang, Jin Tang, and Jianqin Yin.
\newblock Seld-mamba: Selective state-space model for sound event localization and detection with source distance estimation.
\newblock \emph{arXiv preprint arXiv:2408.05057}, 2024.

\bibitem[Chen et~al.(2024)Chen, Yi, Xue, Wang, Zhang, Dong, Zeng, Tao, Zhao, and Fan]{chen2024rawbmamba}
Yujie Chen, Jiangyan Yi, Jun Xue, Chenglong Wang, Xiaohui Zhang, Shunbo Dong, Siding Zeng, Jianhua Tao, Lv~Zhao, and Cunhang Fan.
\newblock Rawbmamba: End-to-end bidirectional state space model for audio deepfake detection.
\newblock \emph{arXiv preprint arXiv:2406.06086}, 2024.

\bibitem[Lin and Hu(2024)]{lin2024audio}
Jiaju Lin and Haoxuan Hu.
\newblock Audio mamba: Pretrained audio state space model for audio tagging.
\newblock \emph{arXiv preprint arXiv:2405.13636}, 2024.

\bibitem[Yadav and Tan(2024)]{yadav2024audio}
Sarthak Yadav and Zheng-Hua Tan.
\newblock Audio mamba: Selective state spaces for self-supervised audio representations.
\newblock \emph{arXiv preprint arXiv:2406.02178}, 2024.

\bibitem[Rafaely(2009)]{rafaely2009spherical}
Boaz Rafaely.
\newblock Spherical loudspeaker array for local active control of sound.
\newblock \emph{The Journal of the Acoustical Society of America}, 125\penalty0 (5):\penalty0 3006--3017, 2009.

\bibitem[Luo et~al.(2024{\natexlab{b}})Luo, Zhou, and Bai]{luo2024mambagan}
Tianhao Luo, Feng Zhou, and Zhongxin Bai.
\newblock Mambagan: Mamba based metric gan for monaural speech enhancement.
\newblock In \emph{2024 International Conference on Asian Language Processing (IALP)}, pages 411--416. IEEE, 2024{\natexlab{b}}.

\bibitem[Shams et~al.(2024)Shams, Dindar, Jiang, and Mesgarani]{shams2024ssamba}
Siavash Shams, Sukru~Samet Dindar, Xilin Jiang, and Nima Mesgarani.
\newblock Ssamba: Self-supervised audio representation learning with mamba state space model.
\newblock \emph{arXiv preprint arXiv:2405.11831}, 2024.

\bibitem[Zhang et~al.(2024{\natexlab{c}})Zhang, Ma, Shahin, Ahmed, and Epps]{zhang2024rethinking}
Xiangyu Zhang, Jianbo Ma, Mostafa Shahin, Beena Ahmed, and Julien Epps.
\newblock Rethinking mamba in speech processing by self-supervised models.
\newblock \emph{arXiv preprint arXiv:2409.07273}, 2024{\natexlab{c}}.

\bibitem[Jiang et~al.(2024{\natexlab{b}})Jiang, Li, Florea, Han, and Mesgarani]{jiang2024speech}
Xilin Jiang, Yinghao~Aaron Li, Adrian~Nicolas Florea, Cong Han, and Nima Mesgarani.
\newblock Speech slytherin: Examining the performance and efficiency of mamba for speech separation, recognition, and synthesis.
\newblock \emph{arXiv preprint arXiv:2407.09732}, 2024{\natexlab{b}}.

\bibitem[Donley et~al.(2017)Donley, Ritz, and Kleijn]{donley2017active}
Jacob Donley, Christian Ritz, and W~Bastiaan Kleijn.
\newblock Active speech control using wave-domain processing with a linear wall of dipole secondary sources.
\newblock In \emph{2017 IEEE International Conference on Acoustics, Speech and Signal Processing (ICASSP)}, pages 456--460. IEEE, 2017.

\bibitem[Kondo and Nakagawa(2007)]{kondo2007speech}
Kazuhiro Kondo and Kiyoshi Nakagawa.
\newblock Speech emission control using active cancellation.
\newblock \emph{Speech communication}, 49\penalty0 (9):\penalty0 687--696, 2007.

\bibitem[Zeghidour et~al.(2021)Zeghidour, Luebs, Omran, Skoglund, and Tagliasacchi]{zeghidour2021soundstream}
Neil Zeghidour, Alejandro Luebs, Ahmed Omran, Jan Skoglund, and Marco Tagliasacchi.
\newblock Soundstream: An end-to-end neural audio codec.
\newblock \emph{IEEE/ACM Transactions on Audio, Speech, and Language Processing}, 30:\penalty0 495--507, 2021.

\bibitem[D{\'e}fossez et~al.(2022)D{\'e}fossez, Copet, Synnaeve, and Adi]{defossez2022high}
Alexandre D{\'e}fossez, Jade Copet, Gabriel Synnaeve, and Yossi Adi.
\newblock High fidelity neural audio compression.
\newblock \emph{arXiv preprint arXiv:2210.13438}, 2022.

\bibitem[Kingma(2014)]{kingma2014adam}
Diederik~P Kingma.
\newblock Adam: A method for stochastic optimization.
\newblock \emph{arXiv preprint arXiv:1412.6980}, 2014.

\bibitem[Luo et~al.(2020)Luo, Chen, and Yoshioka]{luo2020dual}
Yi~Luo, Zhuo Chen, and Takuya Yoshioka.
\newblock Dual-path rnn: efficient long sequence modeling for time-domain single-channel speech separation.
\newblock In \emph{ICASSP 2020-2020 IEEE International Conference on Acoustics, Speech and Signal Processing (ICASSP)}, pages 46--50. IEEE, 2020.

\bibitem[Garofolo(1993)]{garofolo1993timit}
John~S Garofolo.
\newblock Timit acoustic phonetic continuous speech corpus.
\newblock \emph{Linguistic Data Consortium, 1993}, 1993.

\bibitem[Varga and Steeneken(1993)]{varga1993assessment}
Andrew Varga and Herman~JM Steeneken.
\newblock Assessment for automatic speech recognition: Ii. noisex-92: A database and an experiment to study the effect of additive noise on speech recognition systems.
\newblock \emph{Speech communication}, 12\penalty0 (3):\penalty0 247--251, 1993.

\bibitem[Vaswani(2017)]{vaswani2017attention}
A~Vaswani.
\newblock Attention is all you need.
\newblock \emph{Advances in Neural Information Processing Systems}, 2017.

\bibitem[Burgess(1981)]{burgess1981active}
John~C Burgess.
\newblock Active adaptive sound control in a duct: A computer simulation.
\newblock \emph{The Journal of the Acoustical Society of America}, 70\penalty0 (3):\penalty0 715--726, 1981.

\bibitem[Boucher et~al.(1991)Boucher, Elliott, and Nelson]{boucher1991effect}
CC~Boucher, SJ~Elliott, and PA~Nelson.
\newblock Effect of errors in the plant model on the performance of algorithms for adaptive feedforward control.
\newblock In \emph{IEE Proceedings F (Radar and Signal Processing)}, volume 138, pages 313--319. IET, 1991.

\bibitem[Das and Panda(2004)]{das2004active}
Debi~Prasad Das and Ganapati Panda.
\newblock Active mitigation of nonlinear noise processes using a novel filtered-s lms algorithm.
\newblock \emph{IEEE Transactions on Speech and Audio Processing}, 12\penalty0 (3):\penalty0 313--322, 2004.

\bibitem[Kuo and Wu(2005)]{kuo2005nonlinear}
Sen~M Kuo and Hsien-Tsai Wu.
\newblock Nonlinear adaptive bilinear filters for active noise control systems.
\newblock \emph{IEEE Transactions on Circuits and Systems I: Regular Papers}, 52\penalty0 (3):\penalty0 617--624, 2005.

\bibitem[Subakan et~al.(2021{\natexlab{b}})Subakan, Ravanelli, Cornell, Bronzi, and Zhong]{subakan2021attention}
Cem Subakan, Mirco Ravanelli, Samuele Cornell, Mirko Bronzi, and Jianyuan Zhong.
\newblock Attention is all you need in speech separation.
\newblock In \emph{ICASSP 2021-2021 IEEE International Conference on Acoustics, Speech and Signal Processing (ICASSP)}, pages 21--25. IEEE, 2021{\natexlab{b}}.

\bibitem[Tobias and Seara(2005)]{tobias2005leaky}
Orlando~Jos{\'e} Tobias and Rui Seara.
\newblock Leaky-fxlms algorithm: Stochastic analysis for gaussian data and secondary path modeling error.
\newblock \emph{IEEE Transactions on speech and audio processing}, 13\penalty0 (6):\penalty0 1217--1230, 2005.

\bibitem[Patra et~al.(1999)Patra, Pal, Chatterji, and Panda]{patra1999identification}
Jagdish~Chandra Patra, Ranendra~N Pal, BN~Chatterji, and Ganapati Panda.
\newblock Identification of nonlinear dynamic systems using functional link artificial neural networks.
\newblock \emph{IEEE transactions on systems, man, and cybernetics, part b (cybernetics)}, 29\penalty0 (2):\penalty0 254--262, 1999.

\bibitem[Tan and Jiang(2001)]{tan2001adaptive}
Li~Tan and Jean Jiang.
\newblock Adaptive volterra filters for active control of nonlinear noise processes.
\newblock \emph{IEEE Transactions on signal processing}, 49\penalty0 (8):\penalty0 1667--1676, 2001.

\bibitem[Habets et~al.(2008)Habets, Gannot, and Cohen]{habets2008robust}
Emanu{\"e}l~AP Habets, Sharon Gannot, and Israel Cohen.
\newblock Robust early echo cancellation and late echo suppression in the stft domain.
\newblock \emph{Proc. of 11th Int. Worksh. on Acoust. Echo and Noise Control IWAENC 2008}, 2008.

\bibitem[Benesty et~al.(2001)Benesty, G{\"a}nsler, Morgan, Sondhi, Gay, et~al.]{benesty2001advances}
Jacob Benesty, Tomas G{\"a}nsler, Dennis~R Morgan, M~Mohan Sondhi, Steven~L Gay, et~al.
\newblock Advances in network and acoustic echo cancellation.
\newblock 2001.

\bibitem[Ivry et~al.(2021)Ivry, Cohen, and Berdugo]{ivry2021nonlinear}
Amir Ivry, Israel Cohen, and Baruch Berdugo.
\newblock Nonlinear acoustic echo cancellation with deep learning.
\newblock \emph{arXiv preprint arXiv:2106.13754}, 2021.

\bibitem[Gannot and Yeredor(2003)]{gannot2003noise}
Sharon Gannot and Arie Yeredor.
\newblock Noise cancellation with static mixtures of a nonstationary signal and stationary noise.
\newblock \emph{EURASIP Journal on Advances in Signal Processing}, 2002:\penalty0 1--13, 2003.

\bibitem[Oppenheim et~al.(1994)Oppenheim, Weinstein, Zangi, Feder, and Gauger]{oppenheim1994single}
Alan~V Oppenheim, Ehud Weinstein, Kambiz~C Zangi, Meir Feder, and Dan Gauger.
\newblock Single-sensor active noise cancellation.
\newblock \emph{IEEE Transactions on Speech and Audio Processing}, 2\penalty0 (2):\penalty0 285--290, 1994.

\bibitem[Revach et~al.(2021)Revach, Shlezinger, Van~Sloun, and Eldar]{revach2021kalmannet}
Guy Revach, Nir Shlezinger, Ruud~JG Van~Sloun, and Yonina~C Eldar.
\newblock Kalmannet: Data-driven kalman filtering.
\newblock In \emph{ICASSP 2021-2021 IEEE International Conference on Acoustics, Speech and Signal Processing (ICASSP)}, pages 3905--3909. IEEE, 2021.

\bibitem[Luo et~al.(2024{\natexlab{c}})Luo, Shi, Ji, Shen, and Gan]{luo2024real}
Zhengding Luo, Dongyuan Shi, Junwei Ji, Xiaoyi Shen, and Woon-Seng Gan.
\newblock Real-time implementation and explainable ai analysis of delayless cnn-based selective fixed-filter active noise control.
\newblock \emph{Mechanical Systems and Signal Processing}, 214:\penalty0 111364, 2024{\natexlab{c}}.

\bibitem[Luo et~al.(2023{\natexlab{a}})Luo, Shi, Shen, Ji, and Gan]{luo2023gfanc}
Zhengding Luo, Dongyuan Shi, Xiaoyi Shen, Junwei Ji, and Woon-Seng Gan.
\newblock Gfanc-kalman: Generative fixed-filter active noise control with cnn-kalman filtering.
\newblock \emph{IEEE Signal Processing Letters}, 2023{\natexlab{a}}.

\bibitem[Tobias and Seara(2006)]{tobias2006lms}
Orlando~Jos{\'e} Tobias and Rui Seara.
\newblock On the lms algorithm with constant and variable leakage factor in a nonlinear environment.
\newblock \emph{IEEE transactions on signal processing}, 54\penalty0 (9):\penalty0 3448--3458, 2006.

\bibitem[Pandey and Wang(2022)]{pandey2022self}
Ashutosh Pandey and DeLiang Wang.
\newblock Self-attending rnn for speech enhancement to improve cross-corpus generalization.
\newblock \emph{IEEE/ACM Transactions on Audio, Speech, and Language Processing}, 30:\penalty0 1374--1385, 2022.

\bibitem[Shi et~al.(2020)Shi, Gan, Lam, and Wen]{shi2020feedforward}
Dongyuan Shi, Woon-Seng Gan, Bhan Lam, and Shulin Wen.
\newblock Feedforward selective fixed-filter active noise control: Algorithm and implementation.
\newblock \emph{IEEE/ACM Transactions on Audio, Speech, and Language Processing}, 28:\penalty0 1479--1492, 2020.

\bibitem[Shi et~al.(2023{\natexlab{b}})Shi, Gan, Lam, Luo, and Shen]{shi2023transferable}
Dongyuan Shi, Woon-Seng Gan, Bhan Lam, Zhengding Luo, and Xiaoyi Shen.
\newblock Transferable latent of cnn-based selective fixed-filter active noise control.
\newblock \emph{IEEE/ACM Transactions on Audio, Speech, and Language Processing}, 31:\penalty0 2910--2921, 2023{\natexlab{b}}.

\bibitem[Luo et~al.(2023{\natexlab{b}})Luo, Shi, Shen, Ji, and Gan]{luo2023deep}
Zhengding Luo, Dongyuan Shi, Xiaoyi Shen, Junwei Ji, and Woon-Seng Gan.
\newblock Deep generative fixed-filter active noise control.
\newblock In \emph{ICASSP 2023-2023 IEEE International Conference on Acoustics, Speech and Signal Processing (ICASSP)}, pages 1--5. IEEE, 2023{\natexlab{b}}.

\bibitem[Luo et~al.(2023{\natexlab{c}})Luo, Shi, Gan, and Huang]{luo2023delayless}
Zhengding Luo, Dongyuan Shi, Woon-Seng Gan, and Qirui Huang.
\newblock Delayless generative fixed-filter active noise control based on deep learning and bayesian filter.
\newblock \emph{IEEE/ACM Transactions on Audio, Speech, and Language Processing}, 2023{\natexlab{c}}.

\bibitem[Shi et~al.(2022{\natexlab{b}})Shi, Huang, Jiang, and Li]{shi2022integration}
Chuang Shi, Mengjie Huang, Huitian Jiang, and Huiyong Li.
\newblock Integration of anomaly machine sound detection into active noise control to shape the residual sound.
\newblock In \emph{ICASSP 2022-2022 IEEE International Conference on Acoustics, Speech and Signal Processing (ICASSP)}, pages 8692--8696. IEEE, 2022{\natexlab{b}}.

\bibitem[Zhu et~al.(2021)Zhu, Xu, Meng, and Luo]{zhu2021new}
Wenzhao Zhu, Bo~Xu, Zong Meng, and Lei Luo.
\newblock A new dropout leaky control strategy for multi-channel narrowband active noise cancellation in irregular reverberation room.
\newblock In \emph{2021 7th International Conference on Computer and Communications (ICCC)}, pages 1773--1777. IEEE, 2021.

\bibitem[Park and Park(2023)]{park2023integrated}
Seunghyun Park and Daejin Park.
\newblock Integrated 3d active noise cancellation simulation and synthesis platform using tcl.
\newblock In \emph{2023 IEEE 16th International Symposium on Embedded Multicore/Many-core Systems-on-Chip (MCSoC)}, pages 111--116. IEEE, 2023.

\bibitem[Panayotov et~al.(2015)Panayotov, Chen, Povey, and Khudanpur]{panayotov2015librispeech}
Vassil Panayotov, Guoguo Chen, Daniel Povey, and Sanjeev Khudanpur.
\newblock Librispeech: an asr corpus based on public domain audio books.
\newblock In \emph{2015 IEEE international conference on acoustics, speech and signal processing (ICASSP)}, pages 5206--5210. IEEE, 2015.

\bibitem[Garofolo et~al.(1993)Garofolo, Graff, Paul, and Pallett]{garofolo1993csr}
John Garofolo, David Graff, Doug Paul, and David Pallett.
\newblock Csr-i (wsj0) complete ldc93s6a.
\newblock \emph{Web Download. Philadelphia: Linguistic Data Consortium}, 83, 1993.

\bibitem[Gemmeke et~al.(2017)Gemmeke, Ellis, Freedman, Jansen, Lawrence, Moore, Plakal, and Ritter]{gemmeke2017audio}
Jort~F Gemmeke, Daniel~PW Ellis, Dylan Freedman, Aren Jansen, Wade Lawrence, R~Channing Moore, Manoj Plakal, and Marvin Ritter.
\newblock Audio set: An ontology and human-labeled dataset for audio events.
\newblock In \emph{2017 IEEE international conference on acoustics, speech and signal processing (ICASSP)}, pages 776--780. IEEE, 2017.

\bibitem[Jiang and Li(2018)]{JIANG2018139}
Jiguang Jiang and Yun Li.
\newblock Review of active noise control techniques with emphasis on sound quality enhancement.
\newblock \emph{Applied Acoustics}, 136:\penalty0 139--148, 2018.
\newblock \doi{https://doi.org/10.1016/j.apacoust.2018.02.021}.
\newblock URL \url{https://www.sciencedirect.com/science/article/pii/S0003682X17307351}.

\bibitem[Liebich et~al.(2018)Liebich, Fabry, Jax, and Vary]{8577985}
Stefan Liebich, Johannes Fabry, Peter Jax, and Peter Vary.
\newblock Signal processing challenges for active noise cancellation headphones.
\newblock In \emph{Speech Communication; 13th ITG-Symposium}, pages 1--5, 2018.

\bibitem[Kuo and Morgan(1999{\natexlab{b}})]{763310}
S.M. Kuo and D.R. Morgan.
\newblock Active noise control: a tutorial review.
\newblock \emph{Proceedings of the IEEE}, 87\penalty0 (6):\penalty0 943--973, 1999{\natexlab{b}}.
\newblock \doi{10.1109/5.763310}.

\bibitem[Kaymak et~al.(2006)Kaymak, Atherton, Rotter, and Millar]{kaymak}
Erkan Kaymak, Mark Atherton, K~Rotter, and B~Millar.
\newblock Active noise control at high frequencies.
\newblock volume~1, 07 2006.

\bibitem[Diederik(2014)]{diederik2014adam}
P~Kingma Diederik.
\newblock Adam: A method for stochastic optimization.
\newblock \emph{(No Title)}, 2014.

\bibitem[Liebich et~al.(2019)Liebich, Fabry, Jax, and Vary]{Liebich2019AcousticPD}
Stefan Liebich, Johannes Fabry, Peter Jax, and Peter Vary.
\newblock Acoustic path database for anc in-ear headphone development.
\newblock 2019.
\newblock URL \url{https://api.semanticscholar.org/CorpusID:204793245}.

\end{thebibliography}

\clearpage
\newpage

\nocite{*}

\end{document}